\documentclass[twocolumn,aps,prx,reprint,longbibliography,superscriptaddress,nofootinbib]{revtex4-2}

\usepackage{graphicx}
\usepackage{dcolumn}
\usepackage{bm}
\usepackage{hyperref}
\hypersetup{
    colorlinks = true,
    linkcolor = blue,
    urlcolor = gray,
    citecolor = magenta
}

\usepackage{amsfonts}
\usepackage{amsmath, amssymb}

\usepackage[table, dvipsnames]{xcolor}
\usepackage{pifont}
\usepackage{placeins}
\usepackage{stackengine}

\usepackage{dblfloatfix}
\usepackage{lipsum}

\usepackage{colortbl}
\usepackage{makecell}

\makeatletter

    \def\CT@@do@color{%
      \global\let\CT@do@color\relax
            \@tempdima\wd\z@
            \advance\@tempdima\@tempdimb
            \advance\@tempdima\@tempdimc
    \advance\@tempdimb\tabcolsep
    \advance\@tempdimc\tabcolsep
    \advance\@tempdima2\tabcolsep
            \kern-\@tempdimb
            \leaders\vrule
                    \hskip\@tempdima\@plus  1fill
            \kern-\@tempdimc
            \hskip-\wd\z@ \@plus -1fill }
    \makeatother

\definecolor{dgray}{gray}{0.9}
\definecolor{mygray}{gray}{0.97}

\usepackage{changepage}
\usepackage{tabularx}
\usepackage{multirow}
\usepackage{array}

\usepackage[]{mdframed}

\newcolumntype{C}{>{\centering\arraybackslash}p{8.3cm}}
\newcolumntype{K}{>{\centering\arraybackslash}p{6.5cm}}

\newcolumntype{A}[1]{>{\centering\arraybackslash\hspace{0pt}}m{#1}}

\usepackage{makecell}
\usepackage{soul}

\newcommand{\KCuCl}{KCuCl$_{3}$}
\newcommand{\TlCuCl}{TlCuCl$_{3}$}
\newcommand{\XCuCl}{{\it X}CuCl$_{3}$}
\newcommand{\abc}{a\frac b 2 \!\frac c 2}

\begin{document}

\preprint{APS/123-QED}

\title{Triplet nodal lines and Chern bands in \XCuCl ({\it X}= K, Tl)}

\author{Charles B. Walker}
\thanks{These authors contributed equally to this work.}

\author{Matthew Stern}
\thanks{These authors contributed equally to this work.}

\author{Judit Romhányi}
\email{jromhany@uci.edu}
\affiliation{Department of Physics and Astronomy, University of California, Irvine, California 92697, USA}
\date{\today}

\begin{abstract}

We investigate the symmetry-enforced line nodes of the triplet excitations of \XCuCl ({\it X}=Tl, K), showing that they are protected by the nonsymmorphic symmetries and are unaffected by the microscopic details, such as interaction and anisotropy strength, as long as the ground state and the symmetry group remain unaltered.
Extending the conventionally used isotropic spin model for \XCuCl, our analysis includes all the symmetry-allowed anisotropies and gives a detailed account of the role they play in the band topology of triplets. We show that the triplet line nodes carry nontrivial Berry phases and compute their $\mathcal{Z}_2$ topological indices. To investigate the effect of breaking the nonsymmorphic symmetry protecting the triplet nodes, we apply a magnetic field tilted away from the high symmetry $(010)$ axis. We find that while the g-tensor anisotropy behaves as a trivial mass gapping out the triplets, exchange anisotropies supply a nontrivial momentum-dependent mass term. Analogous to Haldane's original model, the competition of these mass terms determines the nature of the band topology in \XCuCl. To enable an analytic study of the band topology we derive an effective Dirac Hamiltonian and validate it by computing the band structure and topological indices in the nodal line and gapped phases from the linear bond-wave formalism.

\end{abstract}

\maketitle

\tableofcontents
\clearpage
\section{\label{sec:intro}Introduction}

Searching for materials that display topological band crossings and exploring the ramifications of nontrivial topology on transport properties have been at the forefront of condensed matter research. The notion of point and line degeneracies realizing Dirac and Weyl physics, first discovered in the electronic band structure of semimetals~\cite{Burkov2011a, Burkov2011b, Fang2015, Fang2016, Yang2022}, has swiftly broadened to include bosonic excitations~\cite{Gao2018, Hu2022, Yang2024, McClarty2022}. 

In this context, Dirac~\cite{Fransson2016, Pershoguba2018, Yao2018, McClarty2019, Yuan2020, Kumar2020, Elliot2021} and Weyl~\cite{Li2016weyl, Mook2016, Su2017, Li2017Weyl} points, Chern bands~\cite{Shindou2013, Zhang2013, Mook2014, Chisnell2015, Chen2018, McClarty2018, Chernyshev2016, Chen2021, Zhu2021, Hu2022magnon, Karaki2023},  nodal line degeneracies~\cite{Mook2017, Li2017, Owerre2017,  Owerre2019, Hwang2020, Scheie2022}, and analogs of $\mathcal{Z}_2$ topology~\cite{Kim2016, Zyuzin2016, Kondo2020} have found realizations in magnon band structures, with (non-quantized) thermal Hall~\cite{Katsura2010, Onose2010, Matsumoto2011a, Matsumoto2011b, Ideue2012, Mook2014Hall, Hirschberger2015, Lee2015, Cao2015, Mook2016TH, Tokiwa2016,  Murakami2017,  Ruckriegel2018, Kawano2019, Furukawa2020, Akazawa2020, Zhuo2021, Neumann2022, Takeda2024,  Ma2024} and spin Nernst~\cite{Cheng2016, Zyuzin2016, Meyer2017, Shiomi2017, Ma2021, Zhang2022} effects arising as their hallmark in transport measurement.  

Triplet excitations are particularly interesting because they carry spin degrees of freedom and have a spin gap protecting the ground state in finite fields. Dimerized quantum magnets have been put forward as hosts of triplet bands exhibiting $\mathcal{Z}$~\cite{
Romhanyi2015, McClarty2017,  Malki2017, Malki2019, Anisimov2019, Nawa2019, Sun2021, Bhowmick2021,  Suetsugu2022,    Esaki2024, Esaki2025} and $\mathcal{Z}_2$~\cite{Joshi2019, Thomasen2021, Kondo2019} topology, most of which, however, remain theoretical models without a material realization. 
The scarce collection of quantum magnets demonstrating nontrivial triplet band topology include Ba$_2$CuSi$_2$O$_6$Cl$_2$ in which the triplet excitations provide an analog of the Su-Schrieffer-Heeger model~\cite{Nawa2019}, SrCu$_2$(BO$_3$)$_2$ exhibiting triplet Chern bands and a spin-1 Dirac cone in an applied magnetic field~\cite{Romhanyi2015, McClarty2017, Bhowmick2021, Sun2021, Suetsugu2022}, and \XCuCl, in which an external electric field has recently been proposed to generate triplet Chern bands~\cite{Esaki2024}.

Here, we put \XCuCl forward as a physical realization of topological triplet line nodes. 
A recent surge of interest in these materials has been fueled by the observation of a magnetoelectric effect~\cite{Kimura2017, Kimura2018, Kimura2020} and the emergence of electric polarization~\cite{Kimura2016} induced by the Bose-Einstein condensation of triplets~\cite{Nikuni2000, Oosawa2001, Ruegg2003, Yamada2008}. Despite the low symmetry of the monoclinic lattice, the triplet modes have been characterized by an isotropic spin Hamiltonian~\cite{Cavadini1999,  Cavadini2001, Cavadini2002, Matsumoto2004}. The importance of the anisotropic Dzyaloshinskii-Moriya (DM) interaction to generate a finite Berry curvature for the triplet bands was first discussed by Esaki~\cite{Esaki2024}, suggesting that by breaking the inversion symmetry a finite DM interaction can be induced on the dimer bonds. This intradimer DM coupling enables the mixing of triplets into the dimer singlet ground state, similar to SrCu$_2$(BO$_3$)$_2$, endowing the triplets with a subsequent Berry curvature and producing a finite thermal Hall transport~\cite{Esaki2024}.
A complete anisotropic Hamiltonian for \XCuCl, however, has not been considered, nor the ramifications of nonsymmorphic symmetry elements for the triplet degeneracies and their topological aspects. 
In this paper, we set out to fill this gap, revealing the nontrivial topology of the triplet line nodes, showing their stability as a consequence of nonsymmorphic symmetries, and discussing the arising band topology when the triplets split in an external magnetic field tilted off of the high-symmetry $(010)$ axis. 

We structure our work as follows: Sec~\ref{sec:model} gives a synopsis of the lattice and symmetry properties, followed by the derivation of the symmetry-allowed exchange anisotropy on each previously studied bond, as well as the full forms of the g-tensors for each site in the unit cell. In Sec.~\ref{sec:method}, we use the linear bond-wave formalism to obtain the fully anisotropic triplet Bogoliubov-de Gennes (BdG) Hamiltonian suitable for \XCuCl. To keep our analysis tractable and attain analytical formulas we derive a low-energy effective Hamiltonian using a canonical transformation that separates the $S^y=1,0,-1$ triplet subspaces based on the Zeeman splitting in an applied field along the $(010)$ direction. Sec.~\ref{sec:triplet_line_nodes} is dedicated to the study of the triplet nodal lines and the characterization of their topology both on the basis of the effective model and by solving the BdG Hamiltonian numerically. In Sec.~\ref{sec:Chern} we investigate the outcome of breaking the nonsymmorphic symmetries that protect the nodal lines by tilting the field away from the $(010)$ axis. We give a detailed analysis of the topology of the gapped triplet modes using our effective model and validating its results with the solution of the full BdG Hamiltonian. We find that the triplet topology is determined by the ratio of the exchange anisotropies and the off-diagonal components of the g-tensor.  Sec.~\ref{sec:discussion} gives a brief summary and outlook.

\section{Model}\label{sec:model}

\begin{figure}[h]
\centering
\includegraphics[width=\columnwidth]{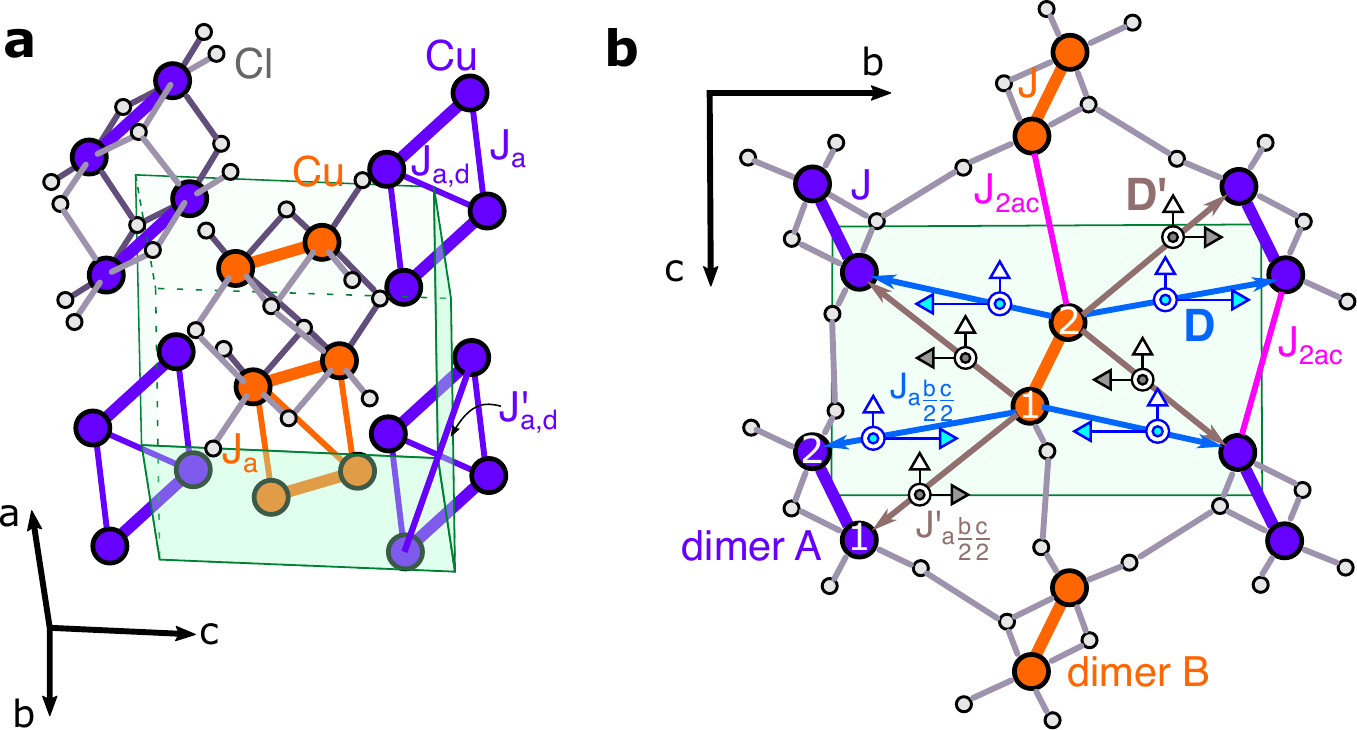}
\caption{(a) Schematic double chain structure of \XCuCl along the a-axis made of CuCl$_6$ complexes. The two types of Cu dimers in the unit cell are denoted with orange and blue colors. (b) Heisenberg and Dzyaloshinskii-Moriya (DM) interactions at the various bonds. The bond arrows indicate the site order considered in the cross product ${\bf S}_i \times{\bf S}_j$ in the DM interaction. The blue and gray bonds couple the different types of dimers along the ${\rm \abc}$-direction. As there are no symmetry elements transforming blue into gray, different interactions along these paths are permitted. The dimers $A$ and $B$, and bonds that connect the same type of dimer, have an inversion center, forbidding DM interactions. Besides the intradimer interactions, we consider interactions along the ${\rm a}$, ${\rm 2ac}$, and ${\rm \abc}$ bonds.}
\label{fig:structure}
\end{figure}

\KCuCl and \TlCuCl crystalize in a monoclinic structure characterized by the space group P2$_1$/c~\cite{Willett1963}. The lattice consists of double dimer chains made of edge-sharing CuCl$_6$ complexes that run along the crystallographic a-axis as shown in Fig.~\ref{fig:structure}. The magnetic copper ions are located in a distorted octahedron of the chlorine ions, resulting in a complete quenching of their orbital degrees of freedom. Magnetization~\cite{Shiramura1997, Oosawa1999, Oosawa2002, Matsumoto2004}, magnetic susceptibility~\cite{Tanaka1996, Takatsu1997}, electron spin resonance~\cite{Kimura2004}, and inelastic neutron scattering~\cite{Kato1998, Kato1999, Cavadini1999, Cavadini2000, Cavadini2001, Cavadini2002, Matsumoto2002, Oosawa2002ins} measurements are consistent with a ground state of a gapped dimer-singlet.  The magnetic properties of \XCuCl have been discussed in terms of a spin-half Heisenberg model, consisting of strong intradimer interactions realized through Cu-Cl-Cu exchanges and weaker interdimer interactions involving two anions via Cu-Cl-Cl-Cu pathways as illustrated in Fig.~\ref{fig:structure}.  The 90$^\circ$ bonds in the CuCl$_6$ planes suggest ferromagnetic interaction, but the small distortion toward a linear bond provides an antiferromagnetic intradimer coupling and realizes a singlet VBC ground state.  

Unlike in SrCu$_2$(BO$_3$)$_2$~\cite{Romhanyi2015, McClarty2017,  Malki2017, Malki2019, Anisimov2019, Nawa2019, Sun2021, Bhowmick2021,  Suetsugu2022,    Esaki2024, Esaki2025} and $\mathcal{Z}_2$~\cite{Joshi2019, Thomasen2021, Kondo2019}, the VBS state in \XCuCl has pure singlet dimers, protected by the inversion symmetry that prevents intradimer Dzyaloshinskii-Moriya (DM) interaction that could mix the triplets into the singlet state. 

Above a critical magnetic field applied in the direction of the b-axis, the triplet excitations undergo a Bose-Einstein condensation (BEC). In the BEC, the singlet and the triplets mix, breaking the inversion symmetry and leading to the emergence of spin-induced electric polarization in \TlCuCl\cite{Kimura2016}. As an inverse mechanism, an applied external electric field can break the inversion symmetry and generate intradimer DM interactions, leading to the mixing of triplets with the singlet ground state and inducing a finite thermal Hall effect~\cite{Esaki2024}.   

Previous studies of \XCuCl consider isotropic interactions on the dimers and between them along the chains (a-direction), as well as the ${\rm 2ac}$ and ${\rm \abc}$ bonds, shown in Fig.~\ref{fig:structure}. Refs.~\cite{Tanaka1998esr} and~\cite{Kimura2004} point out that a DM interaction can be present in \XCuCl, but its effects are first studied in Ref.~\cite{Esaki2024}.  
To get a comprehensive picture of the topological properties of the triplet bands we derive the full symmetry-allowed spin Hamiltonian for the intradimer, a, ${\rm 2ac}$, and ${\rm \abc}$-bonds and systematically narrow down the large parameter space to a few anisotropies without loss of generality. 

The point group of the lattice is isomorphic to C$_{2\text{h}}$, containing the symmetry elements of identity $\{E | {\bf 0}\}$, inversion $\{I | {\bf 0}\}$, glide plane $\{\sigma_{ac} | {\bf t}\}$, and screw axis $\{C_2(b) | {\bf t}\}$, where the fractional lattice-translation ${\bf t}$ is $(0\frac 1 2 \frac 1 2)$. The low symmetry enables a fully anisotropic exchange matrix for the $({\rm \abc})$-bonds (denoted with gray and blue lines in Fig.~\ref{fig:structure}) that connect different dimers because none of the symmetry elements map those bonds back to themselves. The spin Hamiltonian for the ${\rm \abc}$-bonds connecting the opposite sites, e.g. $A_1$ with $B_2$ has the form $\mathcal{H}_{A_1 B_2}={\bf S}_{A_1}^T H_{A_1 B_2}{\bf S}_{B_2}$ with
\begin{eqnarray}
H_{A_1 B_2}\!=\!\left(\begin{smallmatrix}
J^x_{\rm \abc} & D^z_{\rm \abc} \!+ K^z_{\rm \abc} & \!-D^y_{\rm \abc}\!+K^y_{\rm \abc}\\
\!-D^z_{\rm \abc}\!+K^z_{\rm \abc} & J^y_{\rm \abc} & D^x_{\rm \abc}\!+K^x_{\rm \abc}\\
D^y_{\rm \abc} \!+K^y_3 & \!-D^x_{\rm \abc}\!+K^x_{\rm \abc} & J^z_{\rm \abc} 
\end{smallmatrix}\!\right),
\label{eq:A1B2_hamilton}
\end{eqnarray}
where $D^\alpha_{\rm \abc}$ corresponds to the antisymmetric Dzyaloshinskii-Moriya exchange anisotropy components and $K^\alpha_{\rm \abc}$ denotes the components of the symmetric exchange anisotropy ($\alpha=x,y,z$) along the ${\rm \abc}$ bond.
The Hamiltonians of the other three blue bonds (cf. Fig.~\ref{fig:structure}(b)), namely $\mathcal{H}_{\!A_2B_1\!}$, $\mathcal{H}_{\!B_2A_1\!}$, and $\mathcal{H}_{\!B_1A_2\!}$, are related to $\mathcal{H}_{\!A_1B_2\!}$ by the inversion, the screw axis, and the glide plane, respectively. Naturally, $\mathcal{H}_{\!A_2B_1\!}$ has the same form as $\mathcal{H}_{\!A_1B_2\!}$ but in $\mathcal{H}_{\!B_2A_1\!}$ and $\mathcal{H}_{\!B_1A_2\!}$  the x and z components of the symmetric and antisymmetric exchange anisotropies change sign. We show the components of the DM vector in Fig.~\ref{fig:structure}(b). 

For bonds connecting the same sites of different dimers ($A_1$-$B_1$ or $A_2$-$B_2$ indicated by gray in Fig.~\ref{fig:structure}(b)), we have a fully anisotropic Hamiltonian as Eq.~(\ref{eq:A1B2_hamilton}), containing a new set of parameters $J'^\alpha_{\rm \abc}$, $D'^\alpha_{\rm \abc}$, and $K'^\alpha_{\rm \abc}$. 

The bonds that connect (the opposite sites of) the same type of dimer (A-A and B-B paths) have an inversion center. These bonds include the intradimer bonds as well as the interdimer bonds along the $(100)$ and $(201)$ directions shown in Fig.~\ref{fig:structure}. Although the DM interactions are forbidden by the inversion symmetry along these paths, the exchange Hamiltonian contains intradimer ($J^\alpha_0$) and interdimer ($J^\alpha_{\rm a}$ and $J^\alpha_{\rm 2ac}$) XYZ anisotropies, as well as intradimer ($K^\alpha_0$) and interdimer ($K^\alpha_{\rm a}$ and $K^\alpha_{\rm 2ac}$) symmetric exchange anisotropies.

To include all the anisotropies permitted by symmetry, we consider the g-tensor:  
\begin{eqnarray}
\mathcal{H}^{\rm Zee}_{\rm u.c.}\!=\! - {\bf h}({\bf g}_{A_1}{\bf S}_{A_1}\!+\!{\bf g}_{A_2}{\bf S}_{A_2}\!+\!{\bf g}_{B_1}{\bf S}_{B_1}\!+\!{\bf g}_{B_2}{\bf S}_{B_2})
\end{eqnarray}
Due to the low symmetry, all nine components of the g-tensor are allowed and independent. However the symmetry elements map the four sites in the unit cell to each other, fully determining the relationship between the g-tensors on the $A_1$, $A_2$, $B_1$, and $B_2$ sites. We find that the g-tensors on site 1 and site 2 are identical, but the $g_{xy}$, $g_{yx}$, $g_{yz}$, and $g_{zy}$ components have opposite signs on dimer A and B:
\begin{eqnarray}
{\bf g}_{A_i}\!=\!
\begin{pmatrix}
         g_{xx} & g_{xy} & g_{xz}\\
        g_{yx} & g_{yy} & g_{yz}\\
        g_{zx} & g_{zy} & g_{zz}
    \end{pmatrix} \;,\;\;{\bf g}_{B_i}\!=\!
    \begin{pmatrix}
        g_{xx} & \!\!-g_{xy} & g_{xz}\\
        \!\!-g_{yx} & g_{yy} & \!\!-g_{yz}\\
        g_{zx} & \!\!-g_{zy} & g_{zz}
   \end{pmatrix}\;.\nonumber\\
   \label{eq:g-tensor}
\end{eqnarray}
We will see that including the g-tensor anisotropy has important consequences for the band-topology of the magnetic excitations. 

\section{Methods}\label{sec:method}

\subsection{Bond-wave formalism}

Using the bond operator formalism~\cite{Sachdev1990}, we rewrite the spin operators in terms of the singlet and triplet states of the dimers. Here we use the magnetic triplet basis with the quantization axis along the $y$-axis, so that the $\left|t_m\right>_\Lambda$ triplet corresponds to the spin-1 state $\left| S\!=\!1, S^y\!=\!m\right>_\Lambda$, where $m$ can take the values $1,0,-1$ and $\Lambda$ denotes the dimer ($\Lambda=A,B$):
\begin{subequations}
\begin{eqnarray}
&&\left|s\right>_\Lambda= \frac{1}{\sqrt{2}} \left(\left|\uparrow\downarrow\right>_\Lambda-\left|\downarrow\uparrow\right>_\Lambda\right)\\
&&\left|t_1\right>_\Lambda =\frac 1 2 \left(i\left|\uparrow\uparrow\right>_\Lambda - \left|\uparrow\downarrow\right>_\Lambda-\left|\downarrow\uparrow\right>_\Lambda-i\left|\downarrow\downarrow\right>_\Lambda\right)\\
&&\left|t_0\right>_\Lambda = \frac{1}{\sqrt{2}} \left(i \left|\uparrow\downarrow\right>_\Lambda+i \left|\downarrow\uparrow\right>_\Lambda\right)\\
&&\left|t_{-1}\right>_\Lambda =\frac 1 2 \left(i\left|\uparrow\uparrow\right>_\Lambda +\left|\uparrow\downarrow\right>_\Lambda+\left|\downarrow\uparrow\right>_\Lambda-i\left|\downarrow\downarrow\right>_\Lambda\right)
\end{eqnarray}\label{eq:basis}
\end{subequations}
The ground state in a low magnetic field is the singlet product state $\prod_{\text u.c.} \left|s\right>_A\left|s\right>_B$, and the six different triplet operators $t^\dagger_{\Lambda,m}$ with $\Lambda=A, B$ and $m=1,0,-1$ form the excitations. Following the generalized spin-wave formalism, we use the constraint $s^\dagger_\Lambda s^{\phantom{\dagger}}_\Lambda+\sum_m t^\dagger_{\Lambda,m} t^{\phantom{\dagger}}_{\Lambda,m}=1$ to express $s^{(\dagger)}_\Lambda$ as $\sqrt{1-\sum_m t^\dagger_{\Lambda,m} t^{\phantom{\dagger}}_{\Lambda,m}}$ and expand the square root assuming that the number of triplet excitations is small. The resulting linear bond-wave Hamiltonian has the form $\mathcal{H}_{\rm bw}=\mathcal{H}^{(0)}+\mathcal{H}^{(2)}$, where the first term corresponds to the ground state energy and the second is a Bogoliubov--de Gennes Hamiltonian describing the triplet dynamics
\begin{eqnarray}
\mathcal{H}^{(2)}=\sum_k 
\begin{pmatrix}
{\bf t}^\dagger_{\bf k} \\
{\bf t}^{\phantom{\dagger}}_{-{\bf k}}
\end{pmatrix}
\begin{pmatrix}
{\bf M}^{\phantom{\dagger}}_{\bf k} & {\bf N}^{\phantom{\dagger}}_{\bf k}\\
{\bf N}^{*{\phantom{\dagger}}}_{-{\bf k}} & {\bf M}^{*{\phantom{\dagger}}}_{-{\bf k}}
\end{pmatrix}
\begin{pmatrix}
 {\bf t}^{\phantom{\dagger}}_{\bf k}\\
 {\bf t}^{\dagger}_{-{\bf k}}
 \end{pmatrix}\;,
 \label{eq:BdG}
\end{eqnarray}
with ${\bf t}^\dagger_{\bf k}=({\bf t}^\dagger_{{\bf k},A,1}, {\bf t}^\dagger_{{\bf k},B,1},{\bf t}^\dagger_{{\bf k},A,0}, {\bf t}^\dagger_{{\bf k},B,0},{\bf t}^\dagger_{{\bf k},A,-1},{\bf t}^\dagger_{{\bf k},B,-1})$. The $6\times 6$ matrices $M_{\bf k}$ and $N_{\bf k}$ contain only even functions of ${\bf k}$, due to the $\{I|0\}$ symmetry element. 
The full form of the BdG Hamiltonian is included in Appendix~\ref{app:full_Hamiltonian}. 

In the case of dimer-factorized spin-gap states, the ${\bf M}_{\bf k}$ and ${\bf N}_{\bf k}$ matrices are similar, 
and the topology of the triplet bands is not affected by restricting ourselves to the hopping part. The correspondence between the band topology of the full Hamiltonian and of ${\bf M}_{\bf k}$ has been discussed in Ref.~\cite{Nawa2019}. However, when the hopping and pairing matrices are not similar, such a simplification is not possible, and both the band structure and the band topology are affected by neglecting the pairing terms. Examples of this case are provided in Refs.~\cite{McClarty2018} and~\cite{Yan2024}, where certain exchange anisotropies only occur in the pairing part and need to be included on equal footing or perturbatively. 

From now on, we consider only the hopping part but will check our results by solving the full BdG Hamiltonian numerically. 
As a further simplification that leaves the topology unaffected, we take the Heisenberg limit $J^\alpha_\delta\!=\!J_\delta$ ($\alpha=x,y,z$) along each $\delta$ bond including the $\delta=0$ intrabond coupling and the $\delta=(1\frac 1 2 \frac 1 2)$, $(100)$, and $(201)$ interdimer paths.

The hopping matrix ${\bf M}_{\bf k}$ contains the diagonal (isotropic) part of the exchange Hamiltonians (\ref{eq:A1B2_hamilton}), as well as the symmetric and antisymmetric exchange anisotropies and the effect of the external field:
\begin{equation}
{\bf M}_{\bf k}= {\bf M}^{\rm iso}_{\bf k}+{\bf M}^{\rm sym}_{\bf k}+{\bf M}^{\rm antisym}_{\bf k}+{\bf M}^{\rm field}_{\bf k}\;.
\end{equation}
To better understand the different terms, we write the 6-by-6 matrices as a tensor product of the three-dimensional {\it spin} part corresponding to the $S^y=1,0,-1$ triplets and the two-dimensional {\it orbital} part denoting the two different dimers A and B, providing the basis ${\bf t}^\dagger_{\bf k}=({\bf t}^\dagger_{{\bf k},A,1}, {\bf t}^\dagger_{{\bf k},B,1},{\bf t}^\dagger_{{\bf k},A,0}, {\bf t}^\dagger_{{\bf k},B,0},{\bf t}^\dagger_{{\bf k},A,-1},{\bf t}^\dagger_{{\bf k},B,-1})$.
 
\begin{eqnarray}
{\bf M}^{\rm iso}_{\bf k}=J_{\bf k} \;(I_3\otimes I_2) +2 J_3 \gamma^c_{k} \; (I_3\otimes \sigma^x)
\label{eq:M_iso}
\end{eqnarray}
For simplicity, we introduce the shorthand notation $J_{\bf k} = J_0 + J_1 \cos (k_a) - J_2 \cos (2k_a \!+\! k_c)$, where $J_0$ denotes the intradimer interaction, $J_1$ comprises the interactions along the $(100)$ direction $J_1=J_a-\frac 1 2 (J_{a,d}+J_{a,d}')$ (see Fig.~\ref{fig:structure}(a)), and $J_2=\frac{1}{2}J_{\rm 2ac}$ corresponds to the interaction between the same type of dimers along the $(201)$ directions. 

The second term proportional to $J_3$ is the interdimer interaction connecting dimer A to dimer B, thus it contains the $\sigma^x$ Pauli matrix in the orbital subspace. At the first order in the bond-wave expansion, only differences in the interactions on the blue and gray pathways appear, $J_3=\frac{1}{2}(J'_{\rm \abc}-J_{\rm \abc})$. The geometric factor is 
\begin{eqnarray}
\gamma^c_{\bf k}=\cos\left(\!\tfrac{k_b}{2}\!\right)\cos\left(\!\tfrac{2k_a+k_c}{2}\!\right)\;. 
\label{eq:gamma_c}
\end{eqnarray}
The 3-by-3 identity matrix $I_3$ reflects the isotropic nature of these terms for each spin sector, rendering ${\bf M}^{\rm iso}_{\bf k}$ block-diagonal with three identical 2-by-2 blocks for the $S^y=1,0,-1$ spin components. 

The eigenvalues of ${\bf M}^{\rm iso}_{\bf k}$ give two bands that are threefold degenerate everywhere in the Brillouin zone (BZ) originating from the full spin-rotation symmetry. This limit has been extensively studied in the original isotropic models of \XCuCl~\cite{Cavadini2000, Matsumoto2004}.

Including the anisotropy terms containing the off-diagonal part of the spin-spin Hamiltonian breaks the spin-rotation symmetry, mixing the $S^y=1,0,-1$ subspaces and inducing a spin splitting in the triplet bands. 

The symmetric exchange anisotropy has the form
\begin{eqnarray}
{\bf M}^{\rm sym}_{\bf k}\!=&-&K^x_{\bf k}\;(Q^{yz}\otimes \sigma^z) \!+\!K^y_{\bf k} \; (Q^{zx}\otimes I_2)\!+\!K^z_{\bf k} \; (Q^{xy}\otimes \sigma^z)\nonumber\\
 &+&2 K_3^x \; \gamma^s_{\bf k} \; (Q^{yz}\otimes \sigma^x)
 + 2K_3^y \; \gamma^c_{\bf k} \; (Q^{zx}\otimes \sigma^x)\nonumber\\
 &-& 2K_3^z \; \gamma^s_{\bf k} \; (Q^{xy}\otimes \sigma^x)\;,
 \label{eq:M_sym}
 \end{eqnarray}
where the operators $Q^{\alpha\beta}$ denote the spin-quadrupoles $L^\alpha L^\beta + L^\beta L^\alpha$ allowed for the $S\!=\!1$ spin of the triplets. 
Again, we introduce the short notation $K^\alpha_{\bf k}=(-\frac{K^\alpha_0}{2} + K_1^\alpha\cos (k_a) - K_2^\alpha \cos(2k_a\!+\!k_c))$, where $K^\alpha_0$ are the $\alpha=x,y,z$ components of the symmetric intradimer exchange anisotropy, and $K^\alpha_1 =K^\alpha_a- \frac{1}{2}(K^\alpha_{a,d}+K'^\alpha_{a,d})$ and $K^\alpha_2=\frac{1}{2} K^\alpha_{\rm 2ac}$ are the symmetric interdimer exchange anisotropy components along the $(100)$ and $(201)$ directions respectively. These terms connect dimers of the same type (A with A and B with B); therefore the orbital part contains the diagonal matrices $I_2$ and $\sigma^z$, indicating that while $K^y_{\bf k}$ is identical for A and B, $K^x_{\bf k}$ and $K^z_{\bf k}$ have opposite signs on the two different dimers. The last three terms describe the symmetric exchange anisotropy on bonds connecting different dimers, as indicated by the $\sigma^x$ Pauli matrix in the orbital subspace. 
Here too, $K^\alpha_3$ is the difference in the anisotropy on the gray and blue pathways, $\frac{1}{2}(K'{^\alpha}_{\rm \abc}-K^\alpha_{\rm \abc})$
The geometric factor $\gamma^c_{\bf k}$ has been defined in (\ref{eq:gamma_c}), and 
\begin{eqnarray}
\gamma^s_{\bf k}=\sin\left(\!\tfrac{k_b}{2}\!\right)\sin\left(\!\tfrac{2k_a+k_c}{2}\!\right).
\label{eq:gamma_s}
\end{eqnarray}
The antisymmetric exchange anisotropy terms contain the three components of the Dzyaloshinskii-Moriya interaction viable on the $({\rm \abc})$-bonds that connect different dimers and do not have an inversion center:
 \begin{eqnarray}
{\bf M}^{\rm antisym}_{\bf k}&=&-2D_3^x \gamma^c_{\bf k} \; (L^x\otimes \sigma^y) -2D_3^y \gamma^s_{\bf k}\; (L^y\otimes\sigma^y)\nonumber\\ 
&&+ 2D_3^z \gamma^c_{\bf k} \; (L^z\otimes\sigma^y)\;.
\label{eq:M_antisym}
\end{eqnarray}
The orbital part is off-diagonal depicting the connection between different types of dimer and the spin part of the $D^\alpha_3$ component contains the antisymmetric combination $L^\beta L^\gamma-L^\gamma L^\beta=i L^\alpha$ reminiscent of the cross product ${\bf S}_i\times {\bf S}_j$. The strength of the DM interaction is determined by the difference in the DM terms on the gray and blue bonds: $D^\alpha_3=\frac{1}{2}(D'^\alpha_{\rm \abc}-D^\alpha_{\rm \abc})$.

Lastly, the Zeeman term that encompasses the g-tensor anisotropy has the form 
\begin{eqnarray}
{\bf M}^{\rm field}_{\bf k}&&=(h^x g_{xx} + h^z g_{zx}) \; (L^x \otimes I_2) - h^y g_{yy} \; (L^y\otimes I_2) \nonumber\\
&&- (h^x g_{xz} + h^z g_{zz}) \; (L^z\otimes I_2) + h^y g_{yx} \; (L^x\otimes \sigma^z) \nonumber\\
&&- (h^x g_{xy} + h^z g_{zy}) \; (L^y\otimes \sigma^z) - h^y g_{yz} \; (L^z\otimes \sigma^z)\;.\nonumber\\
\label{eq:M_field}
\end{eqnarray}
 The g-tensor components that couple to $L^\alpha\otimes I_2$ are the same for the A and B dimers, while those that occur with $L^\alpha\otimes\sigma^z$ have opposite signs for A and B (cf. Eq.~\ref{eq:g-tensor}).

\subsection{Effective Dirac Hamiltonian}

A fully isotropic Hamiltonian used to reproduce the triplet dynamics~\cite{Cavadini2000} gives two sets of bands, corresponding to the orbital degree of freedom (dimers A and B), each threefold degenerate with the $S^y=1,0,-1$ spin components in the entire BZ as a consequence of the SU(2) spin rotation symmetry. The symmetric and antisymmetric exchange anisotropy terms mix the spin components and generate a band splitting away from the high-symmetry lines.  
In this section, we study the topological properties of the triplet bands near the band touching. Conventionally, such an analysis is done in terms of a two-component Weyl Hamiltonian obtained perturbatively near the degenerate points or lines~\cite{Burkov2011a}.

To keep this work and the roles of various anisotropy terms tractable, we derive an effective two-level Dirac Hamiltonian for each $S^y=1,0,-1$ spin component, perturbatively decoupling the spin subspaces: 
\begin{eqnarray} 
\tilde{\bf M}_{k}=
\begin{pmatrix} 
H_{1,{\bf k}} & 0 & 0\\
0 & H_{0,{\bf k}} & 0\\
0 & 0 & H_{-1,{\bf k}}\;.
\end{pmatrix},\label{eq:hopping0}
\end{eqnarray}
The resulting 2-by-2 Hamiltonians $\mathcal{H}_m$ have the generic form of
\begin{eqnarray}
H_{m,{\bf k}}=E_m({\bf k})+{\bf d}_m({\bf k})\cdot\boldsymbol{\sigma},
\label{eq:Dirac}
\end{eqnarray} 
where $E_m({\bf k})$ is a uniform ${\bf k}$-dependent shift within each $m$ subspace. The vector ${\bf d}_m({\bf k})$ acts as a pseudo-magnetic field in momentum space, encoding the topology of the triplet bands, and the vector $\boldsymbol{\sigma}$ contains the three Pauli matrices $\boldsymbol{\sigma}=(\sigma^x,\sigma^y,\sigma^z)$ that describe the orbital degrees of freedom as they do in (\ref{eq:M_iso})--(\ref{eq:M_field}). 

Without the anisotropic terms, the hopping Hamiltonian is readily in the desired block-diagonal form of (\ref{eq:hopping0}), containing only ${\bf M}_{k}^{\rm iso}$. The anisotropies add off-diagonal terms to (\ref{eq:hopping0}), mixing the $S^y$ components. A canonical transformation that brings the hopping Hamiltonian to the block-diagonal form requires a sizable energy difference between the subspaces that will be decoupled. However, the leading energy in (\ref{eq:hopping0}) is the diagonal intradimer interaction $J_0$, which is the same for all $S^y$ components. Thus we include a magnetic field to generate a Zeeman splitting of the triplets and to have a control parameter that separates the different spin subspaces. Applying a magnetic field along the $(010)$ direction does not lower the space group symmetry (although it breaks the spin-rotation and the time-reversal symmetries) as $h^y$ transforms as the fully symmetric irreducible representation of the point group C$_{2{\text h}}$. 
The details of the canonical transformation are provided in Appendix~\ref{app:canonical_transormation}. Here we only provide the final result for the terms in the Dirac Hamiltonian (\ref{eq:Dirac}).
The energy shift has the form
\begin{eqnarray}
E_m({\bf k})&\!=\!&J_{\bf k}-m \;g_{yy}h^y -\frac{m}{2 g_{yy} h^y} K_{\bf k}^2\nonumber\\
&-&\frac{3 m^2\!-\!2}{g_{yy}}(g_{yz} K^x_{\bf k}  + 
   g_{yx} K^z_{\bf k})\nonumber\\
&-&\frac{m}{2 g_{yy} h^y}\left[
g_{xx}^2 {h^x}^2 + g_{zz}^2 {h^z}^2\right]\nonumber\\
&-&\frac{2m}{g_{yy} h^y}(\gamma^c_{\bf k})^2 \left[\left(K_3^y\right)^2 \!\!+\!\left(D_3^x\right)^2 \!\!+\! \left(D_3^z\right)^2\right]\nonumber\\
&-&\frac{2m}{g_{yy} h^y}(\gamma^s_{\bf k})^2\left[ \left(K_3^x\right)^2 \!\!+\! \left(K_3^z\right)^2\right]
\;,
\end{eqnarray}
where $K_{\bf k}^2=(K^x_{\bf k})^2 + (K^y_{\bf k})^2 + (K^z_{\bf k})^2$.
$J_{\bf k}$ is the uniform energy shift corresponding to the ${\bf k}$-dependent spin gap defined above. The band splittings of the $m=1,0,-1$ triplets due to the Zeeman energy and anisotropies are realized through the terms proportional to $m$ or $3 m^2\!-\!2$. 
The components of the vector ${\bf d}_{m}({\bf k})$ are

\begin{subequations}
\begin{eqnarray}
d^x_m({\bf k})&=&2 J_3 \gamma^c_{\bf k} 
+\frac{2 \!-\! m^2}{g_{yy}}2 \gamma^c_{\bf k} (D_3^z g_{yx} - D_3^x g_{yz}) \nonumber\\
&&-\frac{m}{g_{yy} h^y}
     2\gamma^c_{\bf k} (D_3^x K^x_{\bf k}  +  K_3^y K^y_{\bf k}  - D_3^z K^z_{\bf k})\nonumber\\
&&+\frac{3 m^2 \!-\! 2}{g_{yy} h^y}2 \gamma^s_{\bf k} (K_3^x g_{zz} h^z + K_3^z g_{xx} h^x)\;,
\label{eq:dx}
\end{eqnarray}
\begin{eqnarray}
d^y_m({\bf k})&=&-m \;2D_3^y \gamma^s_{\bf k}
\nonumber\\
&&+\frac{m}{g_{yy}} \;2\gamma^s_{\bf k} \;(K_3^z g_{yz}  \!-\! K_3^x g_{yx}) \nonumber\\
&&-\frac{2 \!-\! m^2}{g_{yy} h^y} \;2\gamma^s_{\bf k} \;(K_3^x K^z_{\bf k} \!-\! K_3^z K^x_{\bf k} ) \nonumber\\
&&+\frac{m}{g_{yy} h^y} \;2\gamma^c_{\bf k} \; (D_3^x g_{xx} h^x + D_3^z g_{zz} h^z)
\;,
\end{eqnarray}
\begin{eqnarray}
d^z_m({\bf k})&=\!&-m \; g_{xy} h^x - m\; g_{zy} h^z\nonumber\\
&&-\frac{m}{g_{yy}} (g_{xx} g_{yx} h^x + g_{yz} g_{zz} h^z)\nonumber\\
&&-\frac{3 m^2 \!-\! 2}{g_{yy} h^y} (g_{zz} h^z K^x_{\bf k} + 
   g_{xx} h^x K^z_{\bf k}) \nonumber\\
&&-\frac{m}{g_{yy} h^y}  \;\gamma^s_{\bf k} \; (D_3^x K_3^x \!-\! D_3^z K_3^z).
\end{eqnarray}
\label{eq:d_vector}
\end{subequations}
For completeness, we included the $x$ and $z$ components of the magnetic field but we work in the case where $\bf{h}$ is very nearly parallel to y, so that these are small perturbations and the decoupling in terms of the $S^y$ component remains valid.

Taking a closer look at ${\bf d}
_m({\bf k})$ allows us to make simplifications to obtain a tractable model without loss of generality. 

The leading term in $d^x_m{{\bf k}}$ is $2 J_3 \gamma^c_{\bf k}$. The next two terms are perturbatively small but are present in a magnetic field parallel to $(010)$. Both depend on ${\bf k}$ through $\gamma^c_{\bf k}$ and thus only renormalize $J_3$ so that it becomes different for the spinful ($m=\pm 1$) and the $m=0$ triplets. The last term has a different ${\bf k}$-dependence but is only present when the field is tilted from $(010)$. Therefore it is a good approximation to just keep the term $2 J_3 \gamma^c_{\bf k}$ for when ${\bf h}\|(010)$.

The leading term in $d^y_m{{\bf k}}$ is $-m \;2D_3^y \gamma^s_{\bf k}$. The next two terms have the same dependence on $\gamma_{\bf k}^s$ and are perturbatively small, renormalizing $D_3^y$. The last term in $d^x_m{{\bf k}}$ depends on $\gamma^c_{\bf k}$ but is only finite when the field is tilted. 

Except for the last, every term in $d^z_m{{\bf k}}$ depends on $h^x$ or $h^z$. The last term is perturbatively small and is only finite if both the DM vector and the anisotropic exchange have $x$ or $z$ components on the $\abc$ bond. 
In the remainder of the paper, we will use the following simplification; we set $D_3^x$, $D_3^z$, and $K_3^\alpha$ ($\alpha=x,y,z$) to zero. In a field parallel to $(010)$, these parameters only show up as a renormalization of $J_3$ in $d^x_m({\bf k})$ and that of $D_3^y$ in $d^y_m({\bf k})$, which does not affect the topology of the bands. And although they generate a finite $d^z_m({\bf k})$ component even without tilting the field, these terms depend on $\gamma^s_{\bf k}$ and so do not split the nodal lines. 

Components of the symmetric exchange anisotropy on the intradimer, $a$, and $2ac$ bonds only appear in the $E_m({\bf k})$ shift that does not affect the topology, as well as in $d^z_m({\bf k})$, coupled to the $x$ and $z$ components of the field. Because the $x$ and $z$ directions are equivalent for spins and only the component of $\bf{K}^x$ (along the $2ac$ bond) which will become important is the component orthogonal to our canting, we will only keep $K_2^x$ and take the $z$ direction as the axis onto which we tilt our $y$ field, which we do without any loss of generality. 
In summary, we retain the Heisenberg exchange interactions $J$, $J_1$, $J_2$, and $J_3$, the $y$ component of the DM interaction $D_3^y$, and the $x$ component of the symmetric exchange anisotropy $K_2^x$ on the $a$ and $2ac$ bonds, along with every element of the g-tensor. 

In what follows, we use the parameters: $J=4.221$ meV, $J_1=-0.212$ meV, $J_2=0.395$ meV, $J_3=0.352$ meV (as taken from Ref.~\cite{Matsumoto2004} for \TlCuCl), $D^y_3=0.2$ meV, $K_2^x=0.2$ meV, $g_{yy}=2.06$, $h^y=18.5$ T, and for tilted fields we set $g_{yx}=g_{yz}=g_{zx}=g_{zy}=0.05$ and $h^z=8.5$ T. The actual values of anisotropies are likely smaller but we decided to use these values to be able to better visualize the gap openings. 

\section{Triplet nodal lines}\label{sec:triplet_line_nodes}

The energy spectrum of the generic Dirac equation~(\ref{eq:Dirac}) has the form 
\begin{eqnarray}
\omega_m({\bf k})=E_m({\bf k}) \pm \;d_m({\bf k}),
\end{eqnarray}
where $d_m({\bf k})=|{\bf d}_m({\bf k})|$ is the length of the pseudo-magnetic field that splits the two levels in each $m$ subspace.

\subsection{Line nodes and band structure for magnetic field along the $(010)$ direction}

\begin{figure}[h]
\centering
\includegraphics[width=\columnwidth]{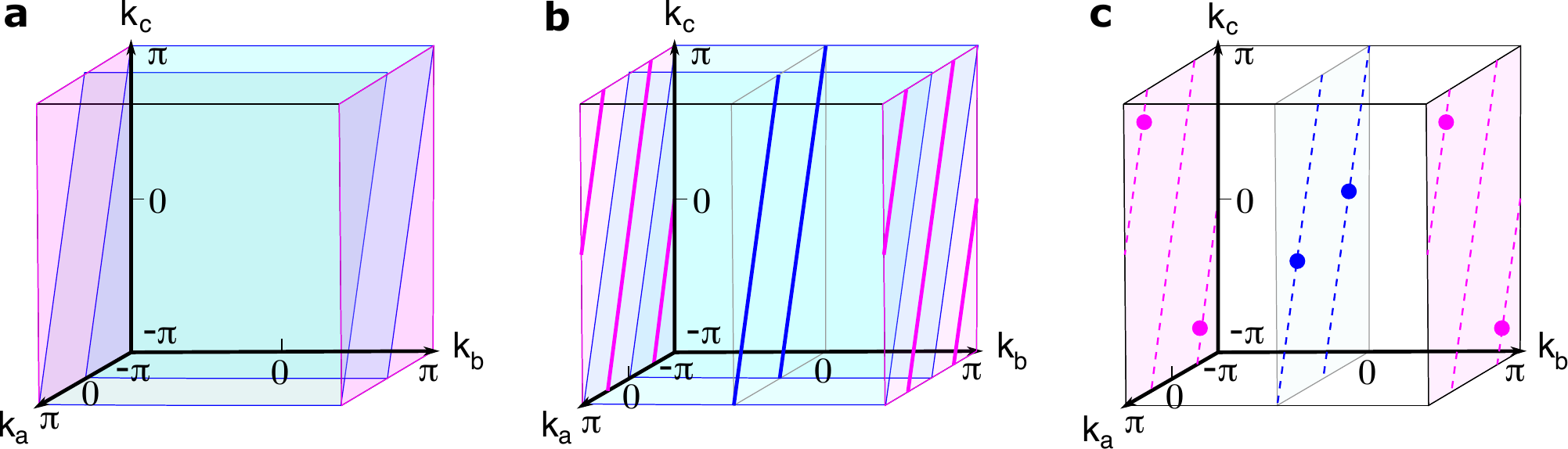}
\caption{(a) Nodal planes at $k_b=(2n_1 +1)\pi$ (magenta) and $k_c=-2 k_a+(2n_2 +1)\pi$ (cyan) in the Heisenberg limit when all the anisotropies are taken to be zero. (b) Including the symmetry-allowed anisotropies reduces the nodal planes to nodal lines. The nodal lines form at $k_b=(2n_1+1)\pi$, $k_c=-2 k_a + 2 n_2 \pi$ (magenta) and at $k_b = 2 n_1 \pi$, $k_c =-2 k_a + (2 n_2+1) \pi$ (blue), with  $n_1,n_2\in \mathbb{Z}$. These nodal lines are protected by the nonsymmorphic symmetries. (c) Tilting the field away from the $(010)$-axis breaks the nonsymmorphic symmetries and splits the nodal line degeneracies that shrink to nodal points. When the anisotropy terms are not finetuned, these degenerate Weyl points split too, leaving fully gapped triplet Chern bands behind.}
\label{fig:nodals}
\end{figure}

In the Heisenberg limit and in zero magnetic field, $E_m({\bf k})=J_{\bf k}$ and ${\bf d}_m({\bf k})$ have only one nonzero component:
\begin{eqnarray}
{\bf d}_m({\bf k})=(
2 J_3 \gamma^c_{\bf k},0,0)\;.\label{eq:d0}
\end{eqnarray}
The single-component ${\bf d}_m({\bf k})$ vector leads to a trivial band structure with degenerate nodal planes at $k_b=(2n_1 +1)\pi$ or $k_c=-2 k_a+(2n_2 +1)\pi$, where $\gamma^c_{\bf k}$ vanishes, as shown in Fig.~\ref{fig:nodals}(a) with magenta and cyan planes respectively. 
The energy of the bands in this isotropic limit is $\omega_m({\bf k})=J_{\bf k}\pm 2 J_3 \gamma^c_{\bf k}$, which is identical to the original result for the triplet bands derived from a Heisenberg model~\cite{Cavadini2000}. Furthermore, $\omega_m({\bf k})$ is independent of $m$ leading to degeneracy in the triplet components. 

The two-fold degeneracy of the nodal planes prevails even in a finite field along the $(010)$ direction, as long as the $g$-tensor anisotropy is neglected. Then the energies become $\omega_m({\bf k})=J_{\bf k}\pm 2 J_3 \gamma^c_{\bf k}-m g_{yy}h^y$ where the last term Zeeman-splits the $m=1,0,-1$ subspaces without affecting the nodal planes of each sector. 

Taking the anisotropies into account gives a more accurate picture of the triplet bands and their topological properties tied to the nonsymmorphic group elements; the $\{\sigma_{ac} | {\bf t}\}$ glide plane and the $\{C_2(b) | {\bf t}\}$ screw axis. Naturally, the anisotropy terms preserve the nonsymmorphic symmetries and thus any degeneracy they enforce. As discussed above, a magnetic field applied in the $(010)$ direction does not reduce the symmetry (as long as we stay in the nonmagnetic dimer singlet ground state), therefore, we consider the fully anisotropic model in an external magnetic field $h^y$. 

Each term in the $x$ and $y$ components of the ${\bf d}_m({\bf k})$ vector in (\ref{eq:d_vector}) depends on ${\bf k}$ via $\gamma^c_{\bf k}$ or $\gamma^s_{\bf k}$. In $d^z_m({\bf k})$ the first four terms vanish when the field is strictly along the $(010)$ direction, and the remaining term has ${\bf k}$ dependence through $\gamma^s_{\bf k}$. Thus for fields ${\bf h}=(0,h^y,0)$, ${\bf d}_m({\bf k})$ becomes zero if both $\gamma^c_{\bf k}$ and $\gamma^s_{\bf k}$ vanish, regardless of the values of the various anisotropies. This condition is fulfilled when $k_b=(2n_1+1)\pi$ and $k_c=-2 k_a + 2 n_2 \pi$, represented by magenta lines, as well as for $k_b = 2 n_1 \pi$ and $k_c =-2 k_a + (2 n_2+1) \pi$, $n_1,n_2\in \mathbb{Z}$, indicated by blue lines in Fig.~\ref{fig:nodals}(b).

Nodal lines, in general, don't enjoy full topological protection but can be stabilized by certain discrete crystallographic symmetries~\cite{Burkov2011a}. In \XCuCl the nonsymmorphic symmetries protect the nodal lines, which is indicated by the vanishing of the ${\bf d}_m({\bf k})$ vector even in the fully anisotropic case. The symmetry protection of the line degeneracy is shown in more detail in the Appendix~\ref{sec:nodal_lines}.

\begin{figure}[hb!]
\centering
\includegraphics[width=0.9\columnwidth]{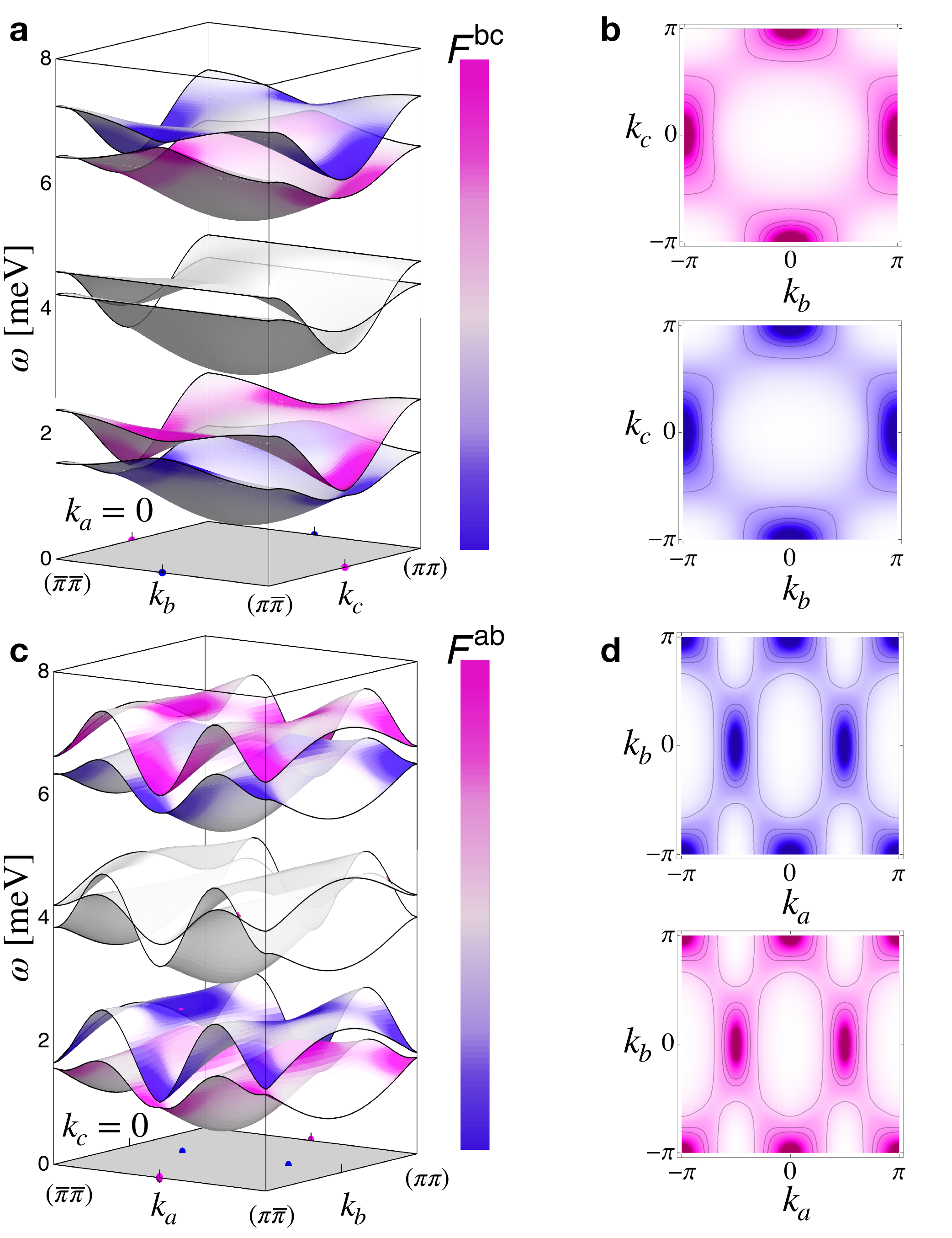}
\caption{Triplet bands in the $ac$-plane for $k_b=\pi$ (a) and $k_b=0$ (b). The nodal lines are indicated with magenta and blue lines following the notation of Fig.~\ref{fig:nodals}. For  $k_b=\pi$ the middle $m=0$ band is degenerate everywhere in the $ac$-plane. Triplet band structure in the $ab$-plane for $k_c=-2k_a+\pi$ (c) and $k_c=-2 k_a$ (d). The nodal lines occur at $k_b = 2 n_1 \pi$, marked with blue in (c) and at $k_b=(2n_1+1)\pi$, denoted with magenta in (d). The $m=0$ band remains degenerate in the $ab$-plane when $k_c=-2k_a+\pi$. The color of the bands encodes the Berry curvature, which in the case of nodal lines is identically zero everywhere where it is well-defined.}
\label{fig:nodal_spectrum}
\end{figure}

For simplicity, we consider only the $y$ component of the DM vector ${\bf D}_3\!=\!(0,D^y_3,0)$ as the leading anisotropy and set each component of the symmetric exchange anisotropies on the various bonds to zero, which does not affect the existence of the nodal lines or their topological properties.  
With this simplification 
\begin{eqnarray}
E_m({\bf k})\!&=&\!J_{\bf k}-m \;g_{yy}h^y 
\nonumber\\
{\bf d}_m({\bf k})&=&(
2 J_3 \gamma^c_{\bf k},\;-m \;2D_3^y \gamma^s_{\bf k},\;0)\;,
\label{eq:Dyhy}
\end{eqnarray}
and the band energies are 
\begin{eqnarray}
\omega_m({\bf k})\!=\!J_{\bf k}\!-\!m g_{yy}h^y 
\!\pm \! 2\sqrt{\!\left(J_3 \gamma^c_{\bf k}\right)^2+\left(m D_3^y \gamma^s_{\bf k}\right)^2},
\end{eqnarray}
where we neglected the perturbatively small term in $E_m({\bf k})$ that only adds a constant shift.  
The nodal line triplet bands are shown in Fig.~\ref{fig:nodal_spectrum}.

\subsection{Topology of the line nodes}
Having only two components in the ${\bf d}_m(\bf k)$ vector allows us to define a topological index for the line touching in terms of a complex order parameter $d=d^x_m({\bf k})+i d^y_m({\bf k})=|d|e^{i  \vartheta}$. The $\vartheta$ phase is well-defined everywhere except at the nodes, thus taking any closed curve in the BZ along which the bands are nondegenerate we get 
\begin{eqnarray}
\oint_C dk^\mu \partial_\mu\vartheta = 2\pi n\;,
\label{eq:winding_number}
\end{eqnarray}
where $n\in\mathbb{Z}$ is the winding number~\cite{Burkov2011a}. When the loop $C$ can be contracted to a point inside the BZ, meaning it does not embrace a singularity, it must have a zero winding number. 

The complex order parameter for the triplet nodal line is $d=2 J_3 \gamma^c_{\bf k}-i 2m \;D_3^y \gamma^s_{\bf k}$, providing a phase angle 
\begin{eqnarray}
  \vartheta=  -i \ln\tfrac{J_3 \gamma^c_{\bf k} - 
   i m D_3^y \gamma^s_{\bf k}}{\sqrt{
  \left(J_3 \gamma^c_{\bf k}\right)^2 + 
   \left(m D_3^y \gamma^s_{\bf k}\right)^2}}\;.
   \label{eq:d_arg}
\end{eqnarray}
Using (\ref{eq:d_arg}), the winding number around the nodal lines can be evaluated analytically and is equal to $n=\pm m\;{\rm sgn}D_3^y$ for the blue/magenta nodes. Thus the topological index for the blue (magenta) nodal lines is identical (opposite) to the $S^y$ spin index of the bands for $D_3^y>0$ (and setting $J_3$ positive).

\begin{figure}[ht!]
\centering
\includegraphics[width=\columnwidth]{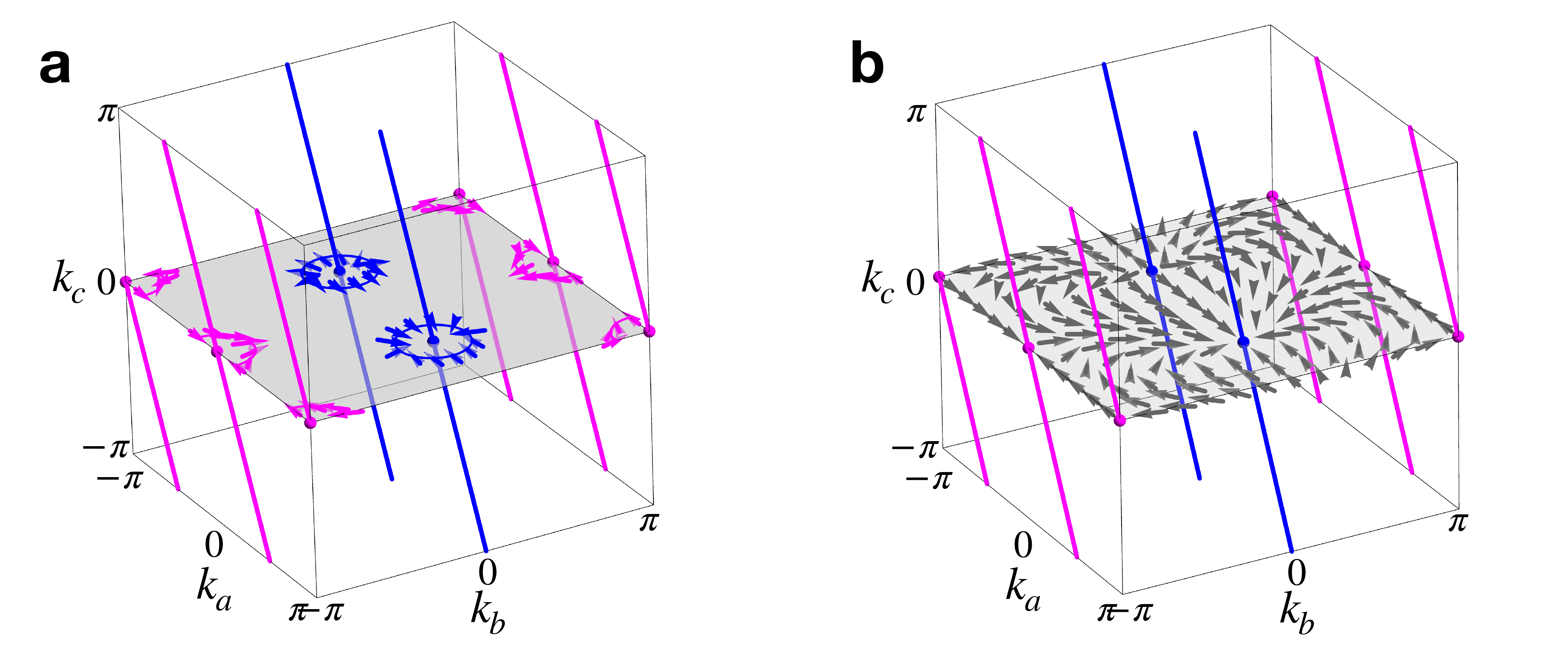}
\caption{(a) Winding of ${\bf d}_m({\bf k})$ for the $m=1$ triplet. ${\bf d}_m({\bf k})$ swirls in an opposite sense around the magenta and blue nodal lines. (b) The behavior of ${\bf d}_m({\bf k})$ the $k_z=0$ plane; the nodal lines behave as vortex lines. A loop that does not enclose them has zero flux and the integral along them gives zero. [Note that we used the lattice gauge here for ${\bf d}_m({\bf k})$, which would not be periodic in the BZ in the physical gauge. ${\bf d}_m({\bf k})=(2 J_3 \gamma^c_{\bf k} \cos\frac{k_b + k_c}{2} - 
  D_3^y \gamma^s_{\bf k}\sin\frac{k_b + k_c}{2}, -D_3^y \gamma^s_{\bf k} \cos\frac{k_b + k_c}{2}+2 J_3 \gamma^c_{\bf k} \sin\frac{k_b + k_c}{2}, 0)$. Changing the gauge does not affect the triplet spectrum or the winding number.]}
\label{fig:d_winding}
\end{figure}

In addition to this analytical result, we may more directly probe the relationship between the nodal lines and our order parameter by expanding around a loop arbitrarily close to each node. 
We begin by investigating the nodal lines at $k_b = 2 n_1 \pi$, $k_c =-2 k_a + (2 n_2+1) \pi$ (blue lines in Fig.~\ref{fig:nodals}). 
As an example, we select the $k_c=0$ plane for the loops and parameterize them as $C_{1/2}$: $k_a=\pm\frac{\pi}{2}+\epsilon \cos\alpha$, $k_b=\epsilon\sin\alpha$ looping around the $(\pm\frac{\pi}{2},0,0)$ points. 
Expanding the complex order parameter $d$ around these points yields $\mp 2 J_3 \epsilon \cos\alpha\mp i  m D_3^y\epsilon\sin\alpha$. For both signs the resulting integral (\ref{eq:winding_number}) is $\int_0^{2\pi}d\alpha\partial_\alpha \vartheta=\int_0^{2\pi}d\alpha\frac{2 m D_3^y J_3}{
4 J_3^2 \cos\alpha^2 + {D_3^y}^2 m^2 \sin\alpha^2}=m\;{\rm sgn}D_3^y$ when $J_3>0$.
Expanding around the magenta nodal lines at the points $(0,\pi,0)$ and $(\pi,\pi,0)$ 
gives $\mp J_3 \epsilon\sin\alpha\mp i m 2 D_3^y \epsilon\cos\alpha $, respectively.
And the winding number becomes $\int_0^{2\pi}d\alpha\partial_\alpha \vartheta=\int_0^{2\pi}d\alpha 
\frac{-2 m D_3^y J_3}{4m^2{D_3^y}^2 \cos\alpha^2 + J_3^2 \sin\alpha^2}=-m\;{\rm sgn}D_3^y$ when $J_3>0$.
The winding of ${\bf d}_m({\bf k})$ is thus opposite for the magenta and blue nodes, as well as for the $1$ and $-1$ triplets.  Having no $y$ component, the $m=0$ triplet is trivial. In Fig.~\ref{fig:d_winding} we show how the ${\bf d}_m({\bf k})$ vector winds around the nodes as well as its behavior in the $k_c=0$ plane.
%
Let us note that the winding number is related to the Berry phase
\begin{equation}
\gamma_m[C]= \oint_C d{\bf k}\cdot \mathcal{A}_m({\bf k})\;,
 \label{eq:Berry_phase}
\end{equation}
where $\mathcal{A}_m({\bf k})=i\langle u_m({\bf k})|\nabla_{\bf k}|u_m({\bf k})\rangle$ is the Berry connection. The eigenstates $|u_m({\bf k})\rangle$ of a two-level Hamiltonian containing only the $x$ and $y$ components of ${\bf d}_m({\bf k})$ are $\frac{1}{\sqrt{2}}\left(\pm\frac{d^x_m({\bf k})-id^y_m({\bf k})}{|{\bf d}_m({\bf k})|},1\right)$, which is equal to $\frac{1}{\sqrt{2}}(\pm e^{i\vartheta},1)$, where $\vartheta$ is the phase angle of $d^x_m({\bf k})+i d^y_m({\bf k})$.  With $|u_m({\bf k})\rangle=(\pm e^{i\vartheta},1)$, the Berry connection is $\mathcal{A}_m({\bf k})=\frac{1}{2}\partial_{\bf k}\vartheta$. Thus, the integral (\ref{eq:Berry_phase}) gives $\pi n$, where $n$ is the winding number. This connection between the winding number and the Berry phase is particularly useful when all three components of ${\bf d}_m({\bf k})$ are finite. In the case of \XCuCl, $d^z_m({\bf k})$ becomes nonzero when $D_3^x$ and $K_3^x$ or $D_3^z$ and $K_3^z$ are included, $d^z_m({\bf k})=\!-\frac{m}{ g_{yy} h^y}  \;\gamma^s_{\bf k} \; (D_3^x K_3^x \!-\! D_3^z K_3^z)$.
The topological index of the nodal line can in general be characterized by the Berry phase formula (\ref{eq:Berry_phase}), which gives $\pm\pi$ for a nontrivial line degeneracy and $0$ for the trivial case. 
The stability of the line nodes can be discussed in terms of a vanishing Berry curvature~\cite{Burkov2011a}, which we computed for the nodal line spectrum and plot in Fig.~\ref{fig:nodal_spectrum} as the color coding of the bands, indicating identically zero Berry curvature everywhere but the nodes.

\section{Triplet Chern bands in a tilted field}\label{sec:Chern}
Next, we consider the effect of tilting the field away from the $(010)$ direction. We chose ${\bf h}=(0,h^y,h^z)$, with $h^z < h^y$ so that the effective two-level models and the separation of the $S^y=m$ subspaces remain valid. The $x$ and $y$ components of ${\bf d}_m({\bf k})$ are unchanged, $2 J_3 \gamma^c_{\bf k}$ and $-m \;2D_3^y \gamma^s_{\bf k}$, respectively, but the $z$ component is now finite
\begin{eqnarray}
d^z_m({\bf k})=\!
- m\; g_{zy} h^z
-m\frac{g_{zz}}{g_{yy}} 
g_{yz}  h^z
-(3 m^2 \!-\! 2)\frac{g_{zz} h^z }{g_{yy} h^y} K^x_{\bf k}.
\nonumber\\
\label{eq:d_z}
\end{eqnarray}
The first two terms in (\ref{eq:d_z}) are present only if the g-tensor anisotropy is fully considered with finite off-diagonal elements. The third term, on the other hand, is finite even without those and instead requires a finite symmetric exchange anisotropy $K^x_{\bf k}=-\frac 1 2 K^x_0 + K^x_1\cos(k_a)-K^x_2\cos(2k_a+k_c)$. 

Tilting the field away from $(010)$ thus introduces a finite $d^z_m({\bf k})$ that acts as a mass term for the nodal lines. [Note that the neglected term $\frac{m}{g_{yy}h^y}\gamma^s_{\bf k} (D^z_3K^z_3-D^x_3K^x_3)$ in $d^z_m({\bf k})$, which does not depend on $h^z$, vanishes at the nodal lines, as these terms are invariant under the nonsymmorphic symmetries protecting the nodes.]  Tilting the field breaks the nonsymmorphic symmetries and opens a gap at the line degeneracies. Similar to Haldane's model, the mass term can open a trivial or a nontrivial gap. 

\subsection{Trivial band gap from g-tensor anisotropy}
\begin{figure}[h!]
\centering
\includegraphics[width=0.9\columnwidth]{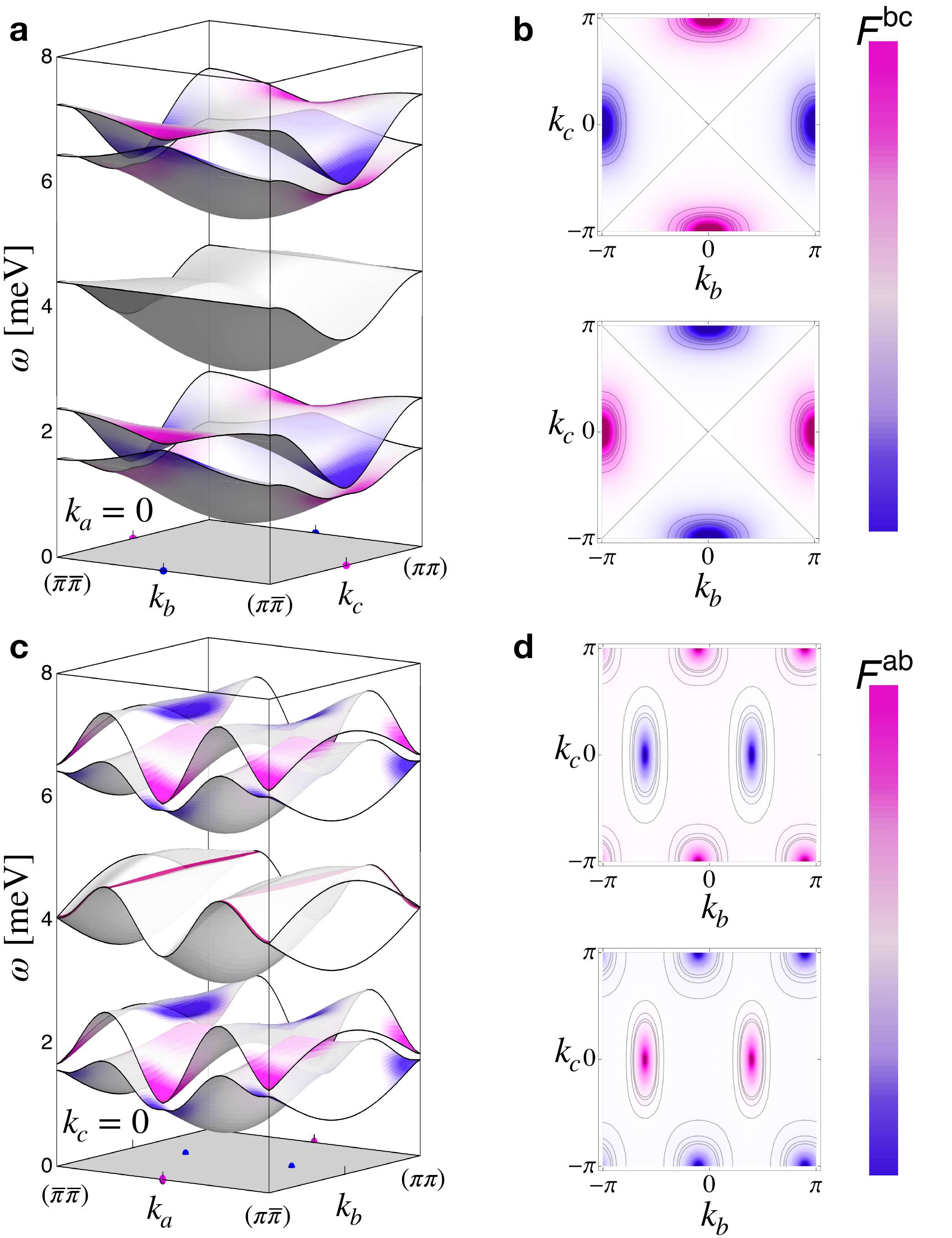}
\caption{Triplet bands in the $bc$-plane for $k_a=0$ (a) and in the $ab$-plane at $k_c=0$ (c), calculated from the ${\bf d}_m({\bf k})$ vector of the effective model. (b) and (d) Density plots of $\frac{1}{2} \hat{{\bf d}}\cdot(\partial_{k_\alpha}\hat{{\bf d}}\times \partial_{k_\beta}\hat{{\bf d}})$. The total Chern number is zero from the cancellation of the Berry curvature.
}
\label{fig:no_Chern}
\end{figure}
When $K^x_{\bf k}=0$, $d^z_m({\bf k})$ corresponds to a ${\bf k}$-independent mass term that has the same sign everywhere in the BZ. Therefore ${\bf d}_m({\bf k})$ does not form a skyrmion in any of the planes in momentum space and all three components of the Chern number ($C^{\alpha\beta}$) vanish due to the correspondence of the Berry curvature and the skyrmion density:
\begin{eqnarray}
C^{\alpha\beta}=\frac{1}{4\pi}\int_{\rm BZ}dk_\alpha dk_\beta \;\hat{{\bf d}}\cdot(\partial_{k_\alpha}\hat{{\bf d}}\times \partial_{k_\beta}\hat{{\bf d}})\;,
\label{eq:Chern}
\end{eqnarray}
where the Berry curvature $F^{\alpha\beta}_{\bf k}$ can be interpreted as the skyrmion density $\frac{1}{2} \hat{{\bf d}}\cdot(\partial_{k_\alpha}\hat{{\bf d}}\times \partial_{k_\beta}\hat{{\bf d}})$, and $\hat{{\bf d}}$ is the ${\bf d}_m({\bf k})$ vector normalized to unity, $\frac{{\bf d}_m({\bf k})}{|{\bf d}_m({\bf k})|}$.

\begin{eqnarray}
{\bf d}_m({\bf k})=(2J_3 \gamma^c_{\bf k}, -2mD^y_3\gamma^s_{\bf k},-2 m g h^z)\;,
\end{eqnarray}
where $g=\frac{1}{2}(g_{zy}+g_{yz}\frac{g_{zz}}{g_{yy}})$ is a constant whose value is close to $g_{zy}$, as it is reasonable to believe that $g_{zz}\approx g_{yy}$ and $g_{zy}\approx g_{yz}$.

The $ac$-plane is trivial for any value of $k_b$ because the cross product $\partial_{k_c}\hat{{\bf d}}\times \partial_{k_a}\hat{{\bf d}}$ vanishes, giving $C^{ac}=0$ $\forall k_b$. This is a consequence of $\gamma^{c}_{\bf k}$ and $\gamma^{s}_{\bf k}$ depending only on the combination of $2k_a+k_c$, rendering $\partial_{k_a}{\bf d}_m({\bf k})=2\partial_{k_c}{\bf d}_m({\bf k})$.

For the $ab$ and $bc$ planes the Berry curvature is finite but the total integral is zero, as expected for a ${\bf d}_m({\bf k})$ vector that does not form a skyrmion in momentum space. We plot the bands in the $ab$ and $bc$ planes in Fig.~\ref{fig:no_Chern} with the Berry curvature distribution indicated on them with a color map. Each region of positive curvature has an associated region of negative curvature, resulting in a zero Chern number.
\begin{figure}[h!]
\centering
\includegraphics[width=\columnwidth]{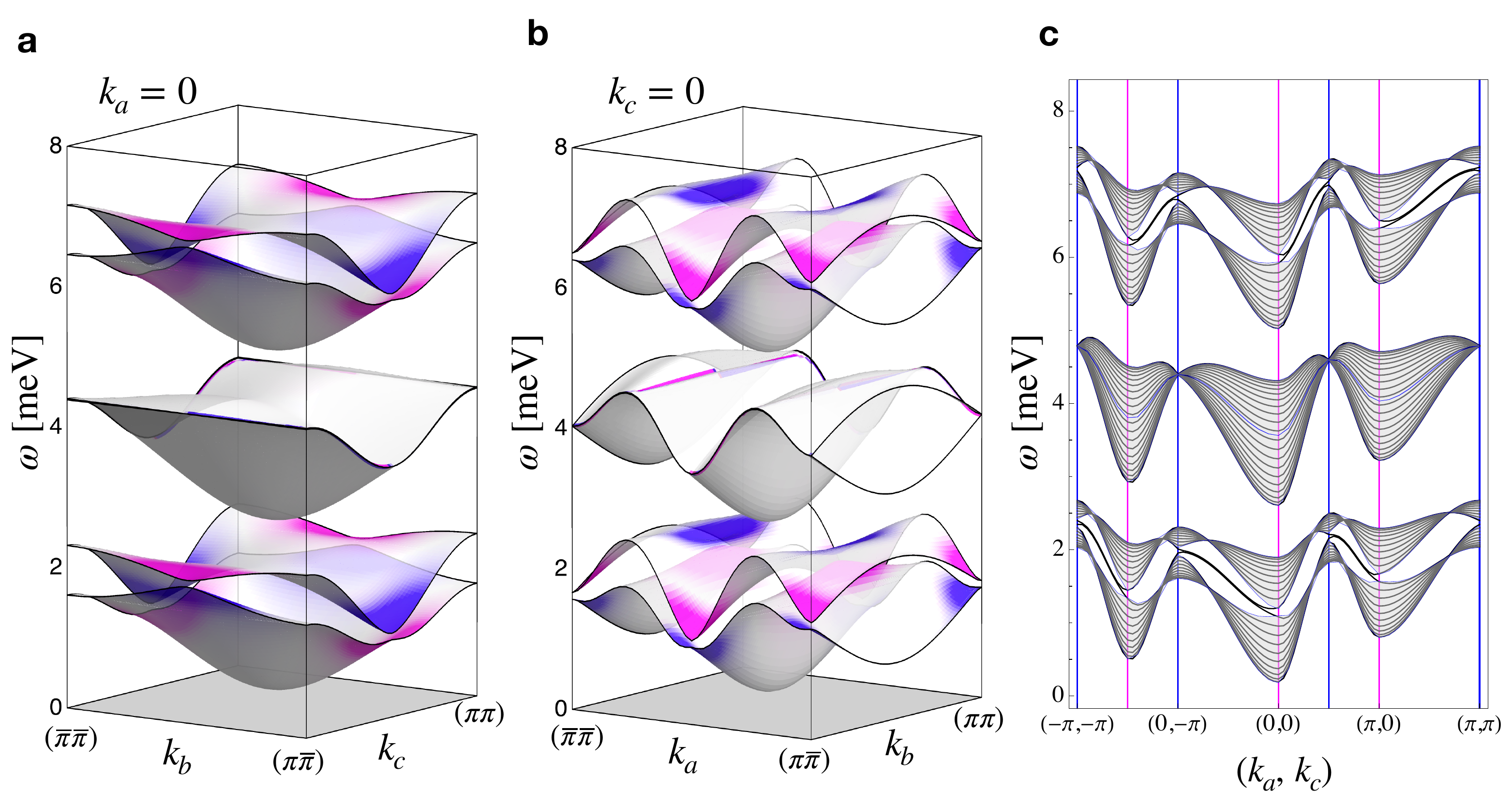}
\caption{Triplet bands in the $bc$-plane for $k_a=0$ (a) and in the $ab$-plane at $k_c=0$ (b), calculated from the original 12-by-12 BdG Hamiltonian without performing canonical transformation. The Berry curvature is obtained numerically, using the method introduced in Ref.~\cite{Fukui2005}. (c) Triplet excitations in open geometry in the $b$ direction are plotted along an irreducible wedge in the BZ. The gray shading is the collapsed band structure of the periodic system. Although there are boundary modes, generic to an open system, they are trivial, corresponding to the vanishing Chern number.
}
\label{fig:no_Chern_12}
\end{figure}
To confirm the validity of our analysis based on the two-level effective model, we numerically computed the triplet spectrum using the full BdG Hamiltonian without relying on perturbation theory. The spectrum of the full bond-wave Hamiltonian is plotted in Fig.~\ref{fig:no_Chern_12}, together with the Berry curvature obtained numerically using the method introduced in Ref.~\cite{Fukui2005}. Furthermore, we calculated the spectrum in open geometry along the $b$ axis using the full BdG Hamiltonian and plotted the triplet excitations along an irreducible wedge in the $ac$ plane. [The shaded gray region corresponds to the collapsed bands along the $k_b$.] The trivial edge modes appearing in the open geometry are a consequence of the vanishing Chern numbers. Both the band structures and Berry curvatures are in excellent agreement with our analytical result from the two-level effective model. 
%

\subsection{Chern bands as an interplay of exchange anisotropy and symmetry-breaking fields }

\begin{figure}[h!]
\centering
\includegraphics[width=0.9\columnwidth]{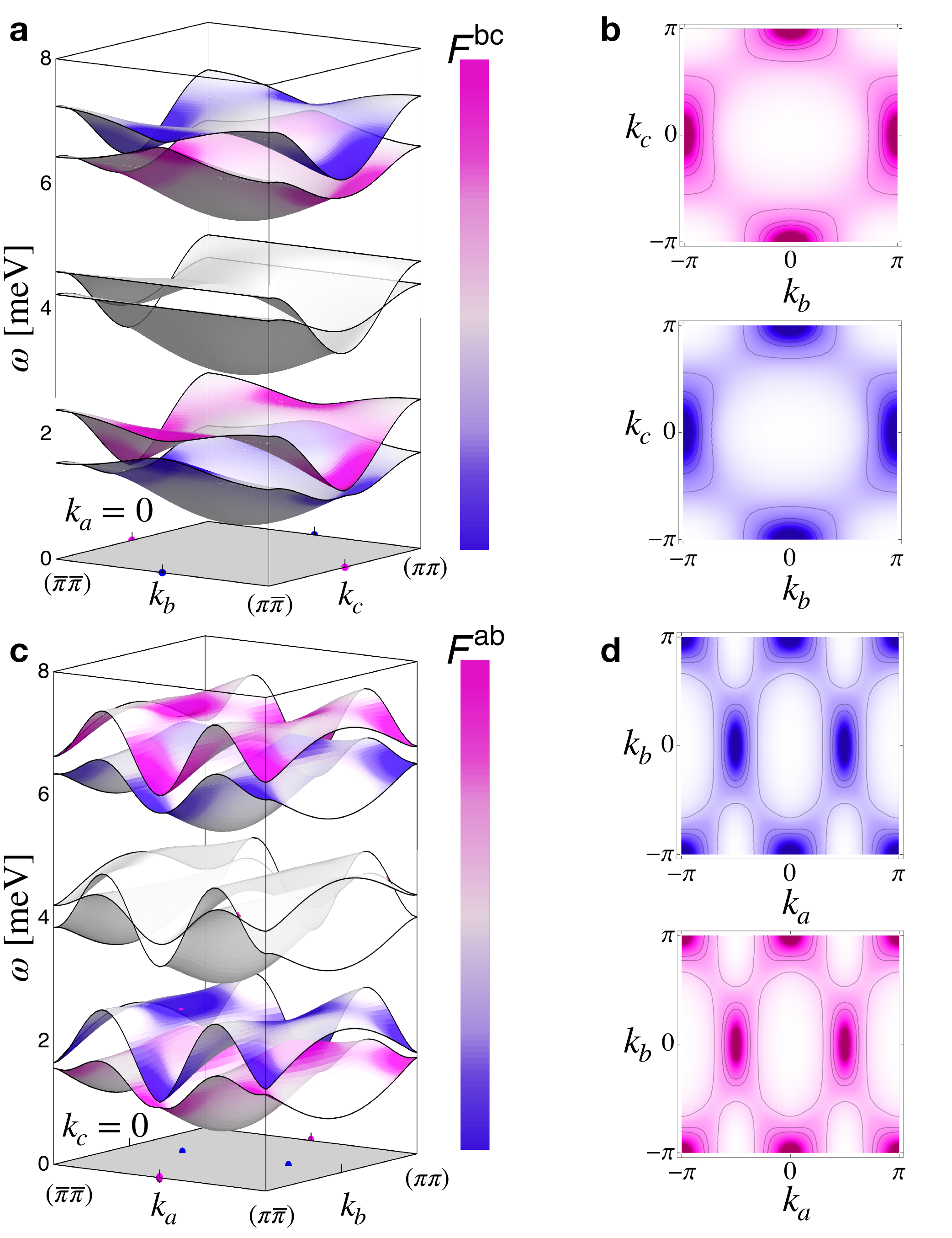}
\caption{Triplet bands in the $bc$-plane for $k_a=0$ (a) and in the $ab$-plane at $k_c=0$ (c), calculated from the ${\bf d}_m({\bf k})$ vector of the effective model. (b) and (d) Density plots of $\frac{1}{2} \hat{{\bf d}}\cdot(\partial_{k_\alpha}\hat{{\bf d}}\times \partial_{k_\beta}\hat{{\bf d}})$. Due to the finite symmetric exchange anisotropy, the total Chern number becomes $\pm m$ for (a) and (b) and $\mp 2m$ for (c) and (d).
}
\label{fig:Chern}
\end{figure}

When $K^x_{\bf k}$ is finite the mass term has a dependence on ${\bf k}$ and a finite Chern number becomes possible depending on the ratio of the trivial and non-trivial mass terms. For simplicity, we set $g_{zz}=g_{yy}$ and $g_{zy}=g_{yz}$, then the trivial mass term is $-2mg_{zy}h^z$ and the non-trivial mass terms become $-(3m^2-2)\frac{h^z}{h^y} K^x_{\bf k}$. Thus, when the off-diagonal g-tensor element, $g_{yz}$, is below the characteristic value $\frac{|K^x_{\bf k}|}{2 h^y}$ value (for $m=\pm 1$), $d^z_m({\bf k})$ can change sign in the BZ, producing a finite skyrmion number.   

We consider only $K^x_{2}$ in $K^x_{\bf k}$ but a similar analysis can be carried out when $K^x_0$, $K^x_1$, or both are kept finite. 
The ${\bf d}_m({\bf k})$ then has the form 
\begin{eqnarray}
{\bf d}_m({\bf k})=(2J_3 \gamma^c_{\bf k}, -2mD^y_3\gamma^s_{\bf k}, \;K^x_{\bf k}\frac{h^z}{h^y}\cos(2 k_x \!+ \!k_z))\;.\nonumber\\
\end{eqnarray}

In the $ac$-plane the Chern number remains zero, as $\partial_{k_c}\hat{{\bf d}}$ and $\partial_{k_a}\hat{{\bf d}}$ are parallel for any $k_b$, rendering the cross product
$\partial_{k_c}\hat{{\bf d}}\times \partial_{k_a}\hat{{\bf d}}=$ 0.
[Note that this is not the case when $K^x_1$ is finite because this term has a different ${\bf k}$-dependence.] 
The Chern numbers in the $ab$ and $bc$ planes can be computed analytically using (\ref{eq:Chern}). In the $bc$ plane at $k_a=0$ the integrand for $m=\pm 1$ is $\frac{m}{4|{\bf d}_m(k_a=0)|^3}D^y_3 J_3 K^x_2\frac{h^z}{h^y} (3 - 2\cos( k_y) \cos (k_z) - \cos(2 k_z))$ which gives $m$, and thus Chern numbers $\pm 1$. This result remains true for any value of $k_a$ because there is no band touching anywhere in the BZ and therefore $C^{bc}$ cannot change as we move along $k_a$.

In the $ab$ plane at $k_c=0$ the integrand in (\ref{eq:Chern}) for $m=\pm 1$ becomes $-\frac{m}{2|{\bf d}_m(k_c=0)|^3}D^y_3 J_3 K^x_2\frac{h^z}{h^y} (3 - 2 \cos(2 k_x) \cos(k_y)-\cos(4 k_x))$ which is the same as the previous skyrmion density aside from the minus sign and the factor 2 multiplying $k_x$, giving a Chern number $-2m$, i.e. $\mp 2$.

\begin{figure}[hb!]
\centering
\includegraphics[width=\columnwidth]{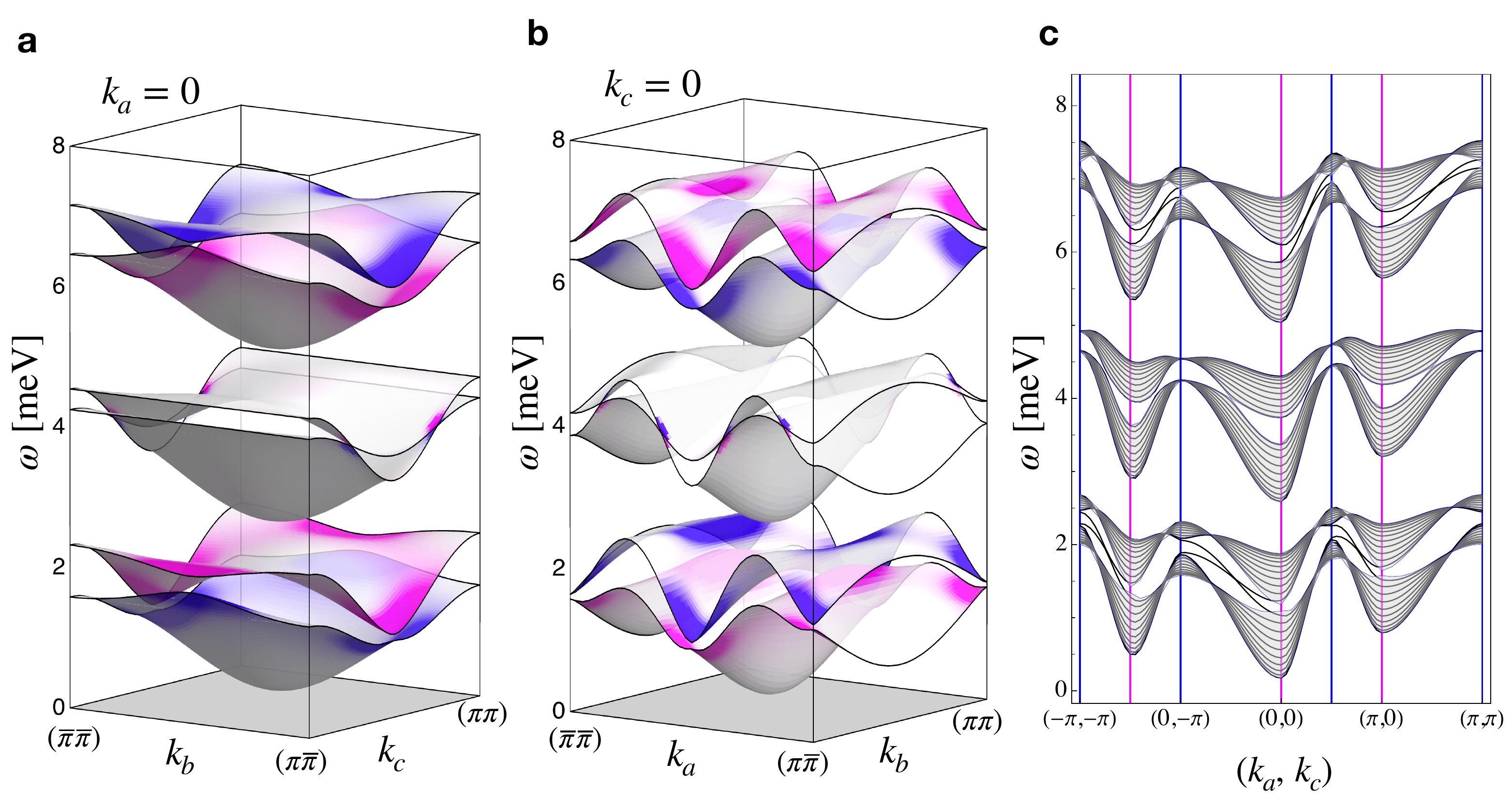}
\caption{Triplet bands and the numerically computed Berry curvature in the $bc$-plane for $k_a=0$ (a) and in the $ab$-plane at $k_c=0$ (b), calculated from the original 12-by-12 BdG Hamiltonian without using perturbation theory. (c) Triplet excitations for open boundary conditions along the $b$ axis plotted on an irreducible wedge in the BZ. The gray shading corresponds to the band structure of the periodic system. We find nontrivial boundary modes corresponding to the finite Chern numbers. 
}
\label{fig:Chern_12}
\end{figure}

We plot the triplet bands and the Berry curvature obtained from the effective two-level model in Fig.~\ref{fig:Chern}. As validation of the effective model, we compute the triplet bands in the periodic and open geometry with termination along the $b$ axis from the original unperturbed BdG Hamiltonian. Additionally, we numerically calculate the Berry curvature of the bands using the technique introduced in~\cite{Fukui2005}. The results of the full triplet BdG Hamiltonian are in excellent agreement with those of our effective model and are shown in Fig.~\ref{fig:Chern_12}.

\section{Conclusion and outlook}\label{sec:discussion}
In addition to symmetries, characterizing physical properties based on topology is gaining increasing importance. Band structures in periodic systems are particularly well-studied in terms of momentum-space topology, traditionally involving a gapped band structure and protected boundary modes. Gapless band structures can also be topologically nontrivial, enabling physical properties that arise from the interplay of bulk states and the characteristic robust surface states.  
The presence of fractional translations combined with symmetry operations in nonsymmorphic systems may result in such band degeneracies, due to their higher-dimensional projective representations at the invariant points, lines, or planes in the Brillouin zone~\cite{bradley2010,Young2015}. Although the nodal line degeneracy is topologically protected, the surface states are generally not, making their decisive experimental detection a challenge.  

Here we propose the dimerized quantum magnet family, \XCuCl ({\it X}=Tl, K), as materials realizing symmetry-protected topological nodal lines in their triplet excitations spectrum.

We establish the full symmetry-allowed exchange Hamiltonian and explore how the band-topology of the triplets is affected by the various anisotropy terms in the dimer-singlet ground state when the magnetic field is large enough to split the triplets but does not yet cause their Bose-Einstein condensation. We show that the triplet line nodes, enforced by the nonsymmorphic symmetries, are topologically nontrivial even without applying an electric field and inducing intradimer DM interaction. They carry a $\pi$ Berry phase and are accompanied by robust boundary modes. 

Upon breaking the nonsymmorphic symmetries, e.g. by an external magnetic field tilted away from the high-symmetry $(010)$ axis, the line degeneracies become gapped. The way in which the topological charge stored in the line nodes spreads to the bands is determined by the relative strength of exchange anisotropies and off-diagonal g-tensor elements. 
The momentum-independent g-tensor anisotropy opens a trivial gap, leading to a canceling Berry curvature distribution on each band. However, interdimer exchange anisotropies realize a nontrivial mass term, splitting the bands so that they each obtain either only positive or negative Berry curvature, producing finite Chern numbers.

These findings are applicable to other dimerized monoclinic systems, such as the isostructural NH$_4$CuCl$_3$ or LiCuCl$_3\cdot 2$H$_2$O and may start a succession of works aimed at identifying experimental fingerprints for symmetry-protected triplet nodal lines. 

Furthermore, the competition of trivial and nontrivial mass terms may have relevance for other quantum magnets. SrCu$_2$(BO$_3$)$_2$, for example, was the first candidate to realize nontrivial triplet bands~\cite{Romhanyi2015, McClarty2017}. The measurement of the triplet thermal Hall effect, however, remains inconclusive~\cite{Cairns2020}. 
The curbed thermal Hall effect in SrCu$_2$(BO$_3$)$_2$ has been argued to originate from the interaction between the triplets~\cite{Suetsugu2022}. Our results suggest the effect of g-tensor anisotropy as an additional factor that can suppress the thermal Hall effect. The symmetry of SrCu$_2$(BO$_3$)$_2$ admits an off-diagonal element of the g-tensor that mixes triplets to the singlet ground state in a different manner as the DM interaction, potentially altering the band topology, calling for further investigations.

\begin{acknowledgments}

This work was supported by the NSF through grant DMR-2142554.
\end{acknowledgments}


\bibliography{KCuCl3}


\clearpage
\appendix 

\onecolumngrid
\section{\label{app:full_Hamiltonian}BdG Hamiltonian}

Here we give the full Bogoliubov-de Gennes Hamiltonian describing the triplet dynamics. 
\begin{eqnarray}
\mathcal{H}^{(2)}=\sum_k 
\begin{pmatrix}
{\bf t}^\dagger_{\bf k} \\
{\bf t}^{\phantom{\dagger}}_{-{\bf k}}
\end{pmatrix}
\begin{pmatrix}
{\bf M}^{\phantom{\dagger}}_{\bf k} & {\bf N}^{\phantom{\dagger}}_{\bf k}\\
{\bf N}^{*{\phantom{\dagger}}}_{-{\bf k}} & {\bf M}^{*{\phantom{\dagger}}}_{-{\bf k}}
\end{pmatrix}
\begin{pmatrix}
 {\bf t}^{\phantom{\dagger}}_{\bf k}\\
 {\bf t}^{\dagger}_{-{\bf k}}
 \end{pmatrix}\;,
 \label{eq:BdG_app}
\end{eqnarray}
where ${\bf t}^\dagger_{\bf k}$ contains the six triplet components with different spin and orbital (i.e. dimer) indices $({\bf t}^\dagger_{{\bf k},A,1}, {\bf t}^\dagger_{{\bf k},B,1},{\bf t}^\dagger_{{\bf k},A,0}, {\bf t}^\dagger_{{\bf k},B,0},{\bf t}^\dagger_{{\bf k},A,-1},{\bf t}^\dagger_{{\bf k},B,-1})$. 
The form of the hopping part ${\bf M}^{\phantom{\dagger}}_{\bf k}$ is discussed in the main text. Here, we give a detailed account of the pairing terms in ${\bf N}^{\phantom{\dagger}}_{\bf k}$.
\begin{eqnarray}
{\bf N}^{\rm iso}_{\bf k}=(J_1 \cos(k_a) - J_2 \cos(2k_a \!+\! k_c)) \; (Q^{y^2}\otimes I_2) +2 J_3 \gamma^c_{\bf k} \; (Q^{y^2}\otimes \sigma_x)
\end{eqnarray}
where $Q^{y^2}=I_3 - 2 (L^y)^2$ is diagonal with the elements $(-1,1,-1)$, rendering ${\bf N}$ the same as ${\bf M}$ for the $S^y=0$ triplet but introducing a minus sign for the $S^y=\pm 1$ triplets. We note that ${\bf N}_{\bf k}$ is independent of the intradimer interaction $J_0$. 
The first term describes the coupling between the same type of dimers and has an identity matrix in its orbital subspace, while the second term originates from the
the interdimer interaction along the $(1\frac 1 2 \frac 1 2)$ path connecting dimer A and dimer B, and thus has $\sigma^x$ in the orbital sector.

\begin{eqnarray}
{\bf N}^{\rm sym}_{\bf k}&=& -\tilde{K}^x \; (i L_x\otimes\sigma_z)
- \tilde{K}^y \; (Q_{zx }\otimes I_2)
-\tilde{K}^z \; (i L_z\otimes\sigma_z)
\nonumber\\
 &+& 2 K^x_3 \gamma^s_{\bf k} \; (i L_x\otimes\sigma_x)
 -2 K^y_3 \gamma^c_{\bf k} \; (Q_{zx} \otimes\sigma_x)
 + 2 K^z_3 \gamma^s_{\bf k} \; (i L_z\otimes\sigma_x)\;,
 \end{eqnarray}
 where $\tilde{K}^\alpha_{\bf k}=K^\alpha_1 \cos(k_a) - K^\alpha_2 \cos(2k_a \!+\! k_c)$ differs from the $K^\alpha_{\bf k}$ in ${\bf M}_{\bf k}$ in that it does not contain the intradimer anisotropy $K^\alpha_0$. [Note that $i L^z= Q^{y^2} Q^{xy}$ and $i L^x= - Q^{y^2} Q^{yz}$, i.e. the x and z components of the symmetric exchange anisotropy for the $S^y=\pm 1$ triplets have opposite signs in ${\bf N}_{\bf k}$ compared to ${\bf M}_{\bf k}$ but the $S^y=0$ triplet terms are the same.]
$K^\alpha_1$ and $K^\alpha_2$ are the symmetric interdimer exchange anisotropy components along the $(100)$ and $(102)$ directions, respectively. These terms connect bonds of the same type and thus have diagonal matrices in the orbital part, with $\sigma_z$ indicating an opposite sign for the A and B dimers. The last three terms describe symmetric exchange anisotropies connecting the A and B dimers, which is signaled by the $\sigma^x$ Pauli matrix in the orbital subspace. The geometric factor $\gamma^c_{\bf k}$ and $\gamma^s_{\bf k}$ are the same as in the main text.

The Dzyaloshinskii-Moriya interaction is only active on the ($3$)-bonds that connect different dimers and have no inversion center:
 \begin{eqnarray}
{\bf N}^{\rm antisym}_{\bf k}=
2 D_3^x \gamma^c_{\bf k} \; (i Q^{yz}\otimes\sigma^y)
+ 2D_3^y \gamma^s_{\bf k} \; (L^y\otimes\sigma^y)
+ 2D_3^z \gamma^c_{\bf k}\; (i Q^{xy}\otimes \sigma^y)
\end{eqnarray}
The off-diagonal orbital part shows the connection between different types of dimers.
[Note that $i Q^{xy}= - Q^{y^2} L^{z}$ and $i Q^{yz}= Q^{y^2} L^{x}$, indicating a sign change for the $S^y=\pm 1$ triplets for ${\bf N}_{\bf k}$.]
Finally, the Zeeman term, similar to the intradimer terms, only appears in ${\bf M}_{\bf k}$. Since $J_0$ and $h^y$ are the leading terms determining the spin gap, we can indeed get away with neglecting the pairing terms. Although one could use a canonical transformation to perturbatively include ${\bf N}$ in a `hopping' Hamiltonian~\cite{Yan2024}, we elected to consider the full BdG Hamiltonian and diagonalize it numerically to check the validity of the approximations described in the main text. 


\section{Effective two-level Hamiltonian from canonical transformation}\label{app:canonical_transormation}

We consider the $M_{\bf k}$ triplet hopping Hamiltonian 
seeking a canonical transformation that eliminates block-off-diagonal terms connecting the $S^y=1,0,-1$ subspaces, such that
	\begin{eqnarray}
		\tilde{\bf M}_{\bf k} 
		=
		\left(\begin{array}{ccc}
			\tilde{\bf H}_{1,\bf k} & {\bm 0} &{\bm 0} \\
            {\bm 0} & \tilde{\bf H}_{0,\bf k} & {\bm 0} \\
			{\bm 0} & {\bm 0} & \tilde{\bf H}_{-1,\bf k} 
		\end{array}\right) \; .
        \label{eq:block-diag_H}
	\end{eqnarray}
In an ideal scenario, this canonical transformation is known exactly and has the form of $e^{-{\bm S}}$ with which
	\begin{eqnarray}
		\tilde{\bf M}_{\bf k} = e^{-{\bm S}} {\bf M}_{\bf k} e^{\bm S} \; ,
	\end{eqnarray} 
	where ${\bm S}$ is an appropriate block--off--diagonal matrix of the same form as (\ref{eq:block-diag_H}).
	When ${\bm S}$ is unknown, we can obtain it perturbatively.
	
	
	Since the block-off-diagonal elements in ${\bm M}_{\bf k}$ are small compared to the diagonal terms, the unitary transformation 	$e^{-{\bm S}}$ must be close to the identity and we can expand the transformation as a power series in ${\bm S}$
	\begin{eqnarray}
		e^{-{\bm S}}= {\bm 1} - {\bm S} +\frac 1 2 {\bm S}^2 -\frac{1}{3!} {\bm S}^3 +\hdots \; .
	\end{eqnarray} 
	The transformed hopping Hamiltonian then becomes
	\begin{equation}
		\tilde{\bf M}_{\bf k} =e^{-{\bm S}} {\bf M}_{\bf k}   e^{{\bm S}} 
		= \sum_n \frac{1}{n!} \left[{\bf M}_{\bf k} , {\bm S} \right]^{(n)} \; , 
	\end{equation}
	with
	\begin{eqnarray}
		[{\bf M}_{\bf k} ,{\bm S}]^{(0)}={\bf M}_{\bf k}  
        \;\; \text{ and}\; \;\;
        \left[ {\bf M}_{\bf k} ,{\bm S} \right]^{(n+1)}= \left[[{\bf M}_{\bf k} ,{\bm S}]^{(n)}, {\bm S} \right]
		  \; .
		\label{eq:canon}
	\end{eqnarray} 
	
	
	In the next step, we separate the block-off-diagonal and block-diagonal parts of $\tilde{\bf M}_{\bf k}$. 
	We decouple the spin-wave Hamiltonian as  
	\begin{eqnarray}
		{\bf M}_{\bf k} = {\bm H}_0+{\bm H}_1+{\bm H}_2 \; ,
	\end{eqnarray} 
	where ${\bm H}_0+{\bm H}_1$ is the block-diagonal part of  ${\bf M}_{\bf k}$, ${\bm H}_0$ containing diagonal elements of ${\bf M}_{\bf k}$ and ${\bm H}_1$ comprising all other block-diagonal terms. 
	Meanwhile, ${\bm H}_2$ is the block-off-diagonal part of ${\bf M}_{\bf k}$, containing the anisotropy terms that mix the $S^y$ subspaces.
	Under the canonical transformation Eq.~(\ref{eq:canon}) these terms transform as  
	\begin{eqnarray}
		\tilde{\bf M}_{{\bf k}, \text{off-diag}} &=&
		\sum_{n=0}^{\infty}\frac{1}{(2n +1)!}\left[{\bm H}_0+{\bm H}_1,{\bm S}\right]^{(2n+1)} 
		+\sum_{n=0}^{\infty}\frac{1}{(2n)!}\left[{\bm H}_2,{\bm S}\right]^{(2n)} \;,
		\label{eq:off_diag}
	\end{eqnarray}
	
	\begin{eqnarray}
		\tilde{\bf M}_{{\bf k},\text{block-diag}} &=& \sum_{n=0}^{\infty}\frac{1}{(2n)!}
		\left[{\bm H}_0 + {\bm H}_1,{\bm S}\right]^{(2n)}
		+\sum_{n=0}^{\infty}\frac{1}{(2n+1)!}\left[{\bm H}_2,{\bm S}\right]^{(2n+1)} \; .
		\label{eq:diag}
	\end{eqnarray} 
	The solution for ${\bm S}$ is provided by the constraint of $\tilde{\bm H}_{\text{off-diag}}=0$. 
	Not knowing the explicit form of the canonical transformation, we expand ${\bm S}$ as a power series 
	\begin{equation}
		{\bm S}={\bm S}_1+{\bm S}_2+{\bm S}_3+\hdots\;,
	\end{equation}
	where ${\bm S}_n$ corresponds to the $n$th order of perturbation. 
	We separate the terms corresponding to different orders in perturbation in $\tilde{\bm H}_{\text{off-diag}}=0$, and obtain the algebraic equations for the matrices ${\bm S}_n$:
	\begin{subequations}
		\begin{eqnarray}
			\left[{\bm H}_0,{\bm S}_1\right]&=&-{\bm H}_2\;,\label{eq:s1}\\
			\left[{\bm H}_0,{\bm S}_2\right]&=&-\left[{\bm H}_1,{\bm S}_1\right]\;,\label{eq:s2}\\
			\left[{\bm H}_0,{\bm S}_3\right]&=&-\left[{\bm H}_1,{\bm S}_2\right]-\frac{1}{3}\left[{\bm H}_2,{\bm S}_1\right]^{(2)}\;,\\
			\vdots\nonumber
		\end{eqnarray}
	\end{subequations}
	Starting with ${\bm S}_1$, we can consecutively solve for the next ${\bm S}_n$ terms. 
	Once we have the ${\bm S}_n$ matrices, we can determine the effective Hamiltonian
	\begin{eqnarray}
		\tilde{\bf M}_{\bf k} = \tilde{\bf M}_{{\bf k},\text{block-diag}}=\sum_{j=0}^\infty \tilde{\bf M}_{\bf k}^{(n)} \; , 
	\end{eqnarray}
	where $\tilde{\bf M}_{\bf k}^{(n)}$ denotes the $n$th order in perturbation 
	
	\begin{subequations}
		\begin{eqnarray}
			\tilde{\bf M}_{\bf k}^{(0)}&=&{\bm H}_0\\
			\tilde{\bf M}_{\bf k}^{(1)}&=&{\bm H}_1\\
			\tilde{\bf M}_{\bf k}^{(2)}&=&
			\left[H_2,{\bm S}_1\right]
			+\frac{1}{2}\left[{\bm H}_0,{\bm S}_1\right]^{(2)}\label{eq:h2}\\
			\tilde{\bf M}_{\bf k}^{(3)}&=&
			\left[H_2,{\bm S}_2\right]
			+\frac{1}{2}\left[{\bm H}_1,{\bm S}_1\right]^{(2)}
			+ \frac{1}{2}\left[\left[{\bm H}_0,{\bm S}_1\right],{\bm S}_2\right]
			+\frac{1}{2}\left[\left[{\bm H}_0,{\bm S}_2\right],{\bm S}_1\right]\\
			\vdots\nonumber
		\end{eqnarray}
		\label{eq:effH_eqs}
	\end{subequations}

\section{\label{sec:nodal_lines}Symmetry Protection of the Nodal Lines}

Our nodal lines in the $h^y \neq 0$, $h^x,h^z = 0$ case fall in the $k_b = 0,\pm \frac{1}{2}$ planes. As the little group of these planes contains only $\{M_{b}|\bf t\}$ and the identity these lines are protected by the nonsymmorphic $\{M_{b}|\bf t\}$ and $\{C_{2b}|\bf t\}$ symmetries of the crystal. Everywhere on this plane our energy eigenstates $|u_m({\bf k})\rangle$ must lie in one of the two one dimensional irreducible representations of the little group $\{ 1,\{M_{b}|\bf t\}\}$ which are distinguished by the values they take $\chi_{G_1}(\{M_{b}|{\bf t}\}) = e^{i\pi k_c}$, $\chi_{G_2}(\{M_{b}|{\bf t}\}) = e^{-i\pi k_c}$ on the nontrivial element. The action of our symmetry on our states is
$$\{M_{b}|{\bf t}\}|u_m({\bf k})\rangle = \lambda_{M,i} e^{-i{\bf t}\cdot {\bf k}} |u_m({\bf k})\rangle$$ where $\lambda_{M,i} = \pm 1$ distinguishes our irreducible representations and $e^{-i{\bf t}\cdot {\bf k}}$ arises due to the translation. We then translate $|u_m({\bf k})\rangle$through momentum space by the reciprocal lattice vector $\hat{k_b}$, which we will call $G$, which preserve our states up to (at most) a complex rotation $ G |u_m({\bf k})\rangle = |u_m({\bf k}+\hat{k_b})\rangle= e^{-i\theta_i}|u_m({\bf k})\rangle$ but composing these operations yields $$\{M_{b}|{\bf t}\}(G |u_m({\bf k})\rangle)= \{M_{b}|{\bf t}\}|u_m({\bf k}+\hat{k_b})\rangle= \lambda_{M,i} e^{-i({\bf t} \cdot \bf k+\pi)} e^{-i\theta_i}|u_m({\bf k})\rangle$$
which exchanges the character of $|u_m({\bf k})\rangle$ under $\{M_{b}|\bf t\}$. This character exchange requires that we pass across a degenerate point, and occurs for any odd $(2n + 1)G, n \in \mathbb{Z}$ translation.

While $G$ is written in terms of a simple translation by $\hat{k_b}$, we could adiabatically evolve $|u_m({\bf k})\rangle$ along any continuous path from ${\bf k}$ to ${\bf k}+\hat{k_b}$ while remaining in the plane. The smallest surface which intersects each path is a nodal line. Simply cutting our plane with an arbitrary line does not necessarily break our $\hat k_b$ loop, but any remaining loops must run parallel to the nodal line. Since a line $k_c = n k_a+c$ experiences $n$ windings around the $k_c$ axis by the time it finishes one loop of the $k_a$ axis, any adiabatic path which returns to the original point must complete a multiple of $n$ windings in the $k_c$ direction. To prohibit all closed loops with odd windings our nodal line must have even integer slope $k_c = 2n k_a+c$ for integer $n$. Additionally, these nodal lines must intersect at least one time reversal invariant momentum (TRIM) point or be doubled by our inversion operator. Such a doubling is not the minimum symmetry protected degeneracy, so our symmetry protected lines must each pass through a TRIM point. This restriction reduces to the requirement that $c = 0,\frac{1}{2}$.

The same argument applies to the four $\{C_{2b}|\bf t\}$ invariant lines, where a single nodal point is sufficient. Our nodal lines provide a good candidate for the source of the nodal points on the $C_{2b}$ invariant lines as these intersection points occur at the TRIM points. In order to minimize our nodal surface, we simply require that the line in the $k_b = 0$ plane and the line in the $k_b = \pm \frac{1}{2}$ plane have different $c = 0, \frac{1}{2}$ so that each $C_{2b}$ invariant line intersects with a nodal line. Because at $\Gamma$ the space group character table of $P2_1/c$ is completely real and one dimensional, the $k_b = 0$ line cannot have $c = 0$, so our nodal lines are ${\bf k} = (x,0,2n x+\frac{1}{2})$ and ${\bf k} = (x,\frac{\pm 1}{2},2n x)$. Determining our last free parameter n is impossible without consulting our actual Hamiltonian.

\end{document}